\documentclass[aps,pra,twocolumn,superscriptaddress,groupedaddress,citeautoscript,preprintnumbers,floatfix]{revtex4}
\pdfoutput=1
\usepackage[latin9]{inputenc}
\usepackage[english]{babel}
\usepackage{graphicx}
\usepackage{bm}
\usepackage{dcolumn} 
\usepackage{amsfonts}
\usepackage{natbib}
\usepackage{upgreek}
\usepackage{amssymb}
\usepackage{amsmath}
\usepackage{physics}
\usepackage{graphicx}
\usepackage{lipsum}
\usepackage{float}
\usepackage[dvipsnames]{xcolor}
\usepackage{url}
\usepackage{hyperref}
\usepackage[normalem]{ulem}
\hypersetup{
    colorlinks=true,
    urlcolor=blue,
    linkcolor=blue,
    citecolor=blue}
\usepackage{upgreek}

\newcommand{\otherlabel}[2]{\protected@edef\@currentlabel{#2}\label{#1}}

\begin{document}

\title{Inertial domain wall characterization in layered multisublattice antiferromagnets}

\author{R. Rama-Eiroa}
\email{ricardo.rama@ehu.eus}
\affiliation{Donostia International Physics Center, 20018 San Sebasti\'an, Spain}
\affiliation{Polymers and Advanced Materials Department: Physics, Chemistry, and Technology, University of the Basque Country, UPV/EHU, 20018 San Sebasti\'an, Spain}
\author{P. E. Roy}
\affiliation{Hitachi Cambridge Laboratory, J. J. Thomson Avenue, Cambridge CB3 0HE, United Kingdom}
\author{J. M. Gonz\'alez}
\affiliation{Polymers and Advanced Materials Department: Physics, Chemistry, and Technology, University of the Basque Country, UPV/EHU, 20018 San Sebasti\'an, Spain}
\author{K. Y. Guslienko}
\affiliation{Polymers and Advanced Materials Department: Physics, Chemistry, and Technology, University of the Basque Country, UPV/EHU, 20018 San Sebasti\'an, Spain}
\affiliation{IKERBASQUE, the Basque Foundation for Science, Plaza Euskadi, 5,
48009 Bilbao, Spain}
\author{J. Wunderlich}
\affiliation{Institute of Physics ASCR, v.v.i., Cukrovarnicka 10, 162 53 Praha 6, Czech Republic}
\affiliation{Institute of Experimental and Applied Physics, University of Regensburg, Universit\"atsstra{\ss}e 31, 93051 Regensburg, Germany}
\author{R. M. Otxoa}
\email{ro274@cam.ac.uk}
\affiliation{Hitachi Cambridge Laboratory, J. J. Thomson Avenue, Cambridge CB3 0HE, United Kingdom}
\affiliation{Donostia International Physics Center, 20018 San Sebasti\'an, Spain}

\date{\today}

\begin{abstract}
The motion of a N\'eel-like ${180}^{\circ}$ domain wall induced by a time-dependent staggered spin-orbit field in the layered collinear antiferromagnet Mn$_2$Au is explored. Through an effective version of the two sublattice nonlinear $\sigma$-model which does not take into account the antiferromagnetic exchange interaction directed along the tetragonal c-axis, it is possible to replicate accurately the relativistic and inertial traces intrinsic to the magnetic texture dynamics obtained through atomistic spin dynamics simulations for quasistatic processes. In the case in which the steady-state magnetic soliton motion is extinguished due to the abrupt shutdown of the external stimulus, its stored relativistic exchange energy is transformed into a complex translational mobility, being the rigid domain wall profile approximation no longer suitable. Although it is not feasible to carry out a detailed follow-up of its temporal evolution in this case, it is possible to predict the inertial-based distance travelled by the domain wall in relation to its steady-state relativistic mass. This exhaustive dynamical characterization for different time-dependent regimes of the driving force is of potential interest in antiferromagnetic domain wall-based device applications.
\end{abstract}

\maketitle

\section{Introduction}\label{sec:intro}

In those spintronics devices that rely on domain walls (DW) as information carriers, the objective is to have ultrafast magnetization dynamics with minimal response times for reasonable external stimuli in order to optimize its operability. Antiferromagnetic (AFM) magnetic solitons can move at velocities of the order of tens of km/s in the special relativity framework \cite{gomonay2016high, shiino2016antiferromagnetic}, and superluminal-like regimes can be accessed for contracted magnetic textures whose extent is comparable to the atomic spacing \cite{yang2019atomic,otxoa2020walker}. However, it is not only important how fast a magnetic texture can move, but also how long it takes for it to move stably at a certain speed. In particular, AFM show a low exchange-mediated static DW mass and a weak motion-based deformation tendency \cite{kim2014propulsion,tatara2020magnon}, therefore it is usually accurate to describe their dynamics through a Newton-like second-order differential equation of motion \cite{tveten2013staggered,yuan2018classification}. This being the case, in the presence of dissipation and external forces the precise magnetic soliton positioning is limited by inertial effects. In this context, it becomes essential to characterize in detail the DW evolution during acceleration and deceleration processes under different time-dependent stimuli. However, in the current literature there is a clear absence of analysis of inertial dynamic signatures of magnetic solitons in real AFM materials with invariant spin spaces where the complete set of interactions are taken into account, as well as certain controversy about the existence of claims about a hypothetical universal AFM DW-like massless behavior \cite{selzer2016inertia}. Among the most interesting systems, it is possible to highlight the case of Mn$_2$Au and CuMnAs, two complex layered AFM that can be excited efficiently through current-induced spin-orbit (SO) fields \cite{vzelezny2014relativistic,wadley2016electrical} and whose magnetic state can be characterized combining magnetoresistance effects with image characterization in real space \cite{olejnik2017antiferromagnetic,grzybowski2017imaging,zhou2018strong,bodnarmagnetoresistance2020}. The experimental observation of DW in these type of materials \cite{barthem2016easy,sapozhnik2018direct,kavspar2021quenching}, as well as proposals based on thermoelectric effects to characterize their positioning \cite{janda2020magneto,otxoa2020giant} and the possibility of extrapolating the usual characterization techniques used in ferromagnets (FM), motivates their theoretical exploration for potential AFM DW-based racetrack memories, all-spintronics architectures, or memristive-like neuromorphic computing approaches \cite{yang2015domain,lequeux2016magnetic,luo2020current}.

In this work, we study the dynamics of a one-dimensional (1D) DW in one of the FM layers of the AFM Mn$_2$Au by means of staggered current-induced SO fields. For this purpose, the crystal and magnetic structure of Mn$_2$Au is introduced in Section \ref{section:system}, as well as all the interactions present in the system. On the other hand, in Section \ref{section:theory} we discuss how it is possible to reduce the description of the system composed of four sublattices to a two sublattice nearest neighbours-based model through the inequivalence in symmetry of the magnetic and crystallographic unit cells. To analyze the magnetic soliton dynamics taking into account the real magnetic structure and interactions of Mn$_2$Au, we introduce an effective version of the nonlinear $\sigma$-model that does not take into account the AFM exchange interaction $\mathcal{J}_2$ along the tetragonal crystal axis, which has a null projection along the DW propagation direction, assumption that is supported by the atomistic spin dynamics simulations shown in Suppl. Notes I and II. Moreover, we demonstrate that it is possible, assuming that the magnetic texture behaves as a rigid entity during its motion, to reduce the Lorentz-invariant formalism to a Newton-like second-order differential equation of motion. To test our theoretical formalism, in Section \ref{section:comparison} we perform atomistic spin dynamics simulations that reveal the relativistic and inertial DW traces for SO field-based quasistatic processes. Notably, when the external stimulus is turned off abruptly interrupting the simulated steady-state motion, the magnetic soliton propagates further than expected via the rigid profile approximation. During the field-free deceleration regime, the relativistic exchange energy stored by the magnetic texture during its previous dynamic evolution is transformed into a complex translational mobility. In this line, we found a reproducible quasilinear correlation between the after-pulse distance travelled by the DW and its steady-state relativistic mass. Finally, the conclusions of our work are exposed in Section \ref{section:conclusions}.

\section{Physical system}\label{section:system}

\begin{figure}[!ht]
\includegraphics[width=8cm]{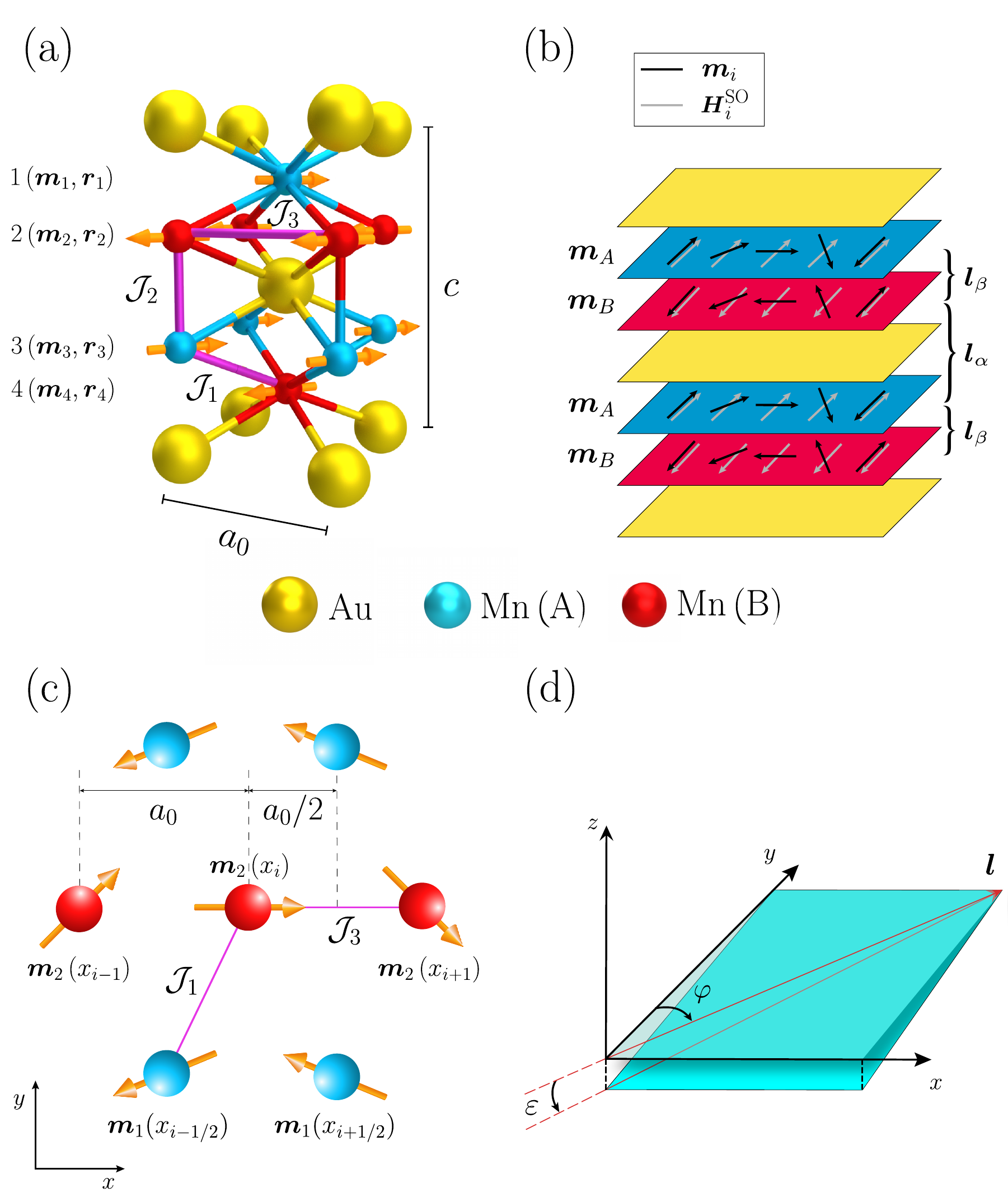}
\caption{(a) Crystal and spin structure of the Mn$_2$Au tetragonal unit cell along with the types of atoms and sublattices present in the system, where the magnetic Mn-based layers are numbered and their corresponding unit magnetization vectors, $\boldsymbol{m}_i$, and position vectors, $\boldsymbol{r}_i$, are indicated. Distribution of the exchange bonds of AFM origin, $\mathcal{J}_1$ and $\mathcal{J}_2$, and the one of FM nature, $\mathcal{J}_3$, as well as the in-plane $xy$ basal lattice period $a_0$ and the out-of-plane height $c$ parameters. (b) Sketch of the distribution of the N\'eel-like DW magnetization, $\boldsymbol{m}_i$, and the SO field, $\boldsymbol{H}^{\mathrm{SO}}_i$, in each of the magnetic sublattices, denoted by A and B, together with the definition of two types of N\'eel order parameters involving different layers, $\boldsymbol{l}_{\alpha}=\left( \boldsymbol{m}_3-\boldsymbol{m}_2 \right)/2$ and $\boldsymbol{l}_{\beta}=\left( \boldsymbol{m}_{1,3}-\boldsymbol{m}_{2,4} \right)/2$. (c) View from the top of the unit cell along the $z$-{\it th} spatial direction of the distribution of the first nearest neighbours of the Mn atom of layer 2 located at the position $x_i$ along the $x$-{\it th} axis, characterized by $\boldsymbol{m}_2 (x_i)$. Those neighbours of its same sublattice, at a distance $a_0$, are denoted by $\boldsymbol{m}_2 (x_{i \pm 1})$, being mediated by the FM exchange interaction $\mathcal{J}_3$, and those from layer 1, located at an in-plane spacing $a_0/2$, are represented by $\boldsymbol{m}_1 (x_{i \pm 1/2})$, and are connected through the AFM exchange contribution encoded by $\mathcal{J}_1$. (d) Description of the unit AFM vector, $\boldsymbol{l}=\left( \boldsymbol{m}_{\mathrm{A}}-\boldsymbol{m}_{\mathrm{B}} \right)/2$, in terms of the polar out-of-plane $\varepsilon$ and in-plane azimuthal $\varphi$ angles relative to the Cartesian coordinate system.}
\label{im:1}
\end{figure}

We consider the layered collinear AFM Mn$_2$Au. This material is interesting because it is a good conductor, it has a strong magnetocrystalline anisotropy \cite{shick2010spin, barthem2013revealing, masrour2015antiferromagnetic}, and a N\'eel temperature well above room temperature \cite{khmelevskyi2008layered}. To characterize the considered system, which is exposed in Fig. \ref{im:1} (a) \cite{wells1970structure}, we write down the interactions that configure the energy, $w$, for the conventional tetragonal unit cell of Mn$_2$Au \cite{roy2016robust, otxoa2020walker}, which is given by
\begin{gather}
w=-\sum_{\left<i,j \right>} \mathcal{J}_{ij} \, \boldsymbol{m}_i \cdot \boldsymbol{m}_j -K_{2 \perp} \sum_i {\left( \boldsymbol{m}_i \cdot \boldsymbol{\hat{z}} \right)}^2 \nonumber \\
- K_{2 \parallel} \sum_i {\left( \boldsymbol{m}_i \cdot \boldsymbol{\hat{y}} \right)}^2-\frac{K_{4 \perp}}{2} \sum_i {\left( \boldsymbol{m}_i \cdot \boldsymbol{\hat{z}} \right)}^4 \nonumber \\
-\frac{K_{4 \parallel}}{2} \sum_i \left[ {\left( \boldsymbol{m}_i \cdot \boldsymbol{\hat{u}}_1 \right)}^4 +{\left( \boldsymbol{m}_i \cdot \boldsymbol{\hat{u}}_2 \right)}^4 \right] \nonumber \\
- \mu_0 \mu_{\mathrm{s}} \sum_i \boldsymbol{m}_i \cdot {\boldsymbol{H}}^{\textrm{SO}}_i,
\label{eq:1}    
\end{gather}
where the sum $\left< i, j \right>$ runs only over first nearest neighbours whose atomic positions are labeled by the indices $i, j$, being represented the unit magnetic moment in the $i$-{\it th} lattice position by $ \boldsymbol{m}_i$. The symbols $\boldsymbol{\hat{x}}$, $\boldsymbol{\hat{y}}$, $\boldsymbol{\hat{z}}$  refer to the unit vectors along the $x$-, $y$-, and $z$-{\it th} spatial directions in the Cartesian coordinate system, while the unit vectors $\boldsymbol{\hat{u}}_{1,2}$ represent the in-plane $xy$-based directions $\boldsymbol{u}_1 = \left[ 1 1 0 \right]$ and $\boldsymbol{u}_2 = \left[ 1 \bar{1} 0 \right]$. Furthermore, as it can be seen in Fig. \ref{im:1} (a), the lattice constant along the $x$- and $y$-{\it th} directions in the basal planes is represented by $a_0=3.328$ \AA \, while the size of the conventional unit cell along the $z$-{\it th} direction is given by $c=8.539$ \AA \, \cite{khmelevskyi2008layered}. Within the conventional unit cell there are two Mn atoms per each type of sublattice, A and B, giving rise to a total of four magnetic atoms. In fact, the magnetic moment of each of these Mn atoms, which coincides with the net contribution to each FM layer in the unit cell, would be given by $\mu_{\mathrm{s}}=4 \mu_{\mathrm{B}}$ \cite{barthem2013revealing}, where $ \mu_{\mathrm{B}} $ is the Bohr magneton. Additionally, as it can be seen in Fig. \ref{im:1} (a), there are three types of exchange contributions between magnetic moments $\boldsymbol{m}_i$ and $\boldsymbol{m}_j$ in the unit cell, which are represented by the exchange integrals $\mathcal{J}_{ij}$. This set is composed by the $\mathcal{J}_1$, $\mathcal{J}_2$, and $\mathcal{J}_3$ parameters, where the first two are AFM, being $\mathcal{J}_1=- \left( 396 \, \mathrm{K} \right) k_{\mathrm{B}}$ and $\mathcal{J}_2=- \left( 532 \, \mathrm{K} \right) k_{\mathrm{B}}$, and the third one is FM, being $\mathcal{J}_3= \left( 115 \, \mathrm{K} \right)  k_{\mathrm{B}}$ \cite{khmelevskyi2008layered, barthem2013revealing, masrour2015antiferromagnetic}, where $k_{\mathrm{B}}$ is the Boltzmann constant. On the other hand, it is possible to appreciate that there are two types of tetragonal anisotropies in the system, given by $K_{4 \parallel}= 1.8548 \times 10^{-25} \, \mathrm{J}$ and $K_{4 \perp}=2 K_{4 \parallel}$, and another two of uniaxial origin, denoted as $K_{2 \perp} = - 1.303 \times 10^{-22} \, \mathrm{J}$ and $K_{2 \parallel} = 7 K_{4 \parallel}$ \cite{shick2010spin,otxoa2020walker,otxoa2020giant}. The last term in Eq. \eqref{eq:1} is the Zeeman-like contribution, where $\mu_0$ denotes the vacuum permeability and $\boldsymbol{H}^{\mathrm{SO}}_i$ expresses the staggered SO field on each $i$-{\it th} lattice site which, because the locally broken inversion symmetry occurs along the $z$-{\it th} spatial direction, is $\boldsymbol{H}^{\mathrm{SO}}_{\mathrm{A}, \mathrm{B}}=\pm H^{\mathrm{SO}}_y \boldsymbol{\hat{y}}$ when the electric current density, $\boldsymbol{j}$, is injected along $\boldsymbol{j} \parallel \boldsymbol{\hat{x}}$, and $\boldsymbol{H}^{\mathrm{SO}}_{\mathrm{A}, \mathrm{B}} = \mp H^{\mathrm{SO}}_x \boldsymbol{\hat{x}}$ when $\boldsymbol{j} \parallel \boldsymbol{\hat{y}}$ \cite{vzelezny2014relativistic}.

\section{Theoretical framework}\label{section:theory}

In view of the different interactions present in the system, collected by Eq. \eqref{eq:1}, the magnetization is constrained in-plane for each Mn-based $xy$ FM layer. Thus, the type of magnetic texture that can be stabilized in each of these planes is a 1D $180^{\circ}$ N\'eel-like DW, as it is exposed in Fig. \ref{im:1} (b). When proposing how it is possible to approach the analytical characterization of the dynamics of a 1D DW in Mn$_2$Au, it is worth noting that the conventional unit cell consists of four staggered magnetized layers along the $z$-{\it th} spatial direction connected through two types of AFM exchange contributions, $\mathcal{J}_1 $ and $ \mathcal{J}_2$, which makes it difficult to define a unique N\'eel order parameter in the system. However, it is feasible to reduce its characterization to a two sublattice single staggered vector-based description due to the symmetric inequivalence of the magnetic and crystallographic unit cells. To carry out this discussion, let us use the numbering of the Mn planes in accordance with Fig. \ref{im:1} (a). This being the case, it is possible to differentiate two crystallographically identical Mn-based groups: one made up of planes 1 and 4, and the other by sheets 2 and 3. We note that, if an inversion transformation is carried out with respect to the unit cell center position, operation which would be characterized through their position vectors $\boldsymbol{r}_i$, it is possible to obtain that crystallographically the Mn atoms of plane 1 are transformed into those of the layer 4, and vice versa (this is, $\boldsymbol{r}_{1,4} \rightarrow - \boldsymbol{r}_{4,1}$). This can be extrapolated to the case of those Mn atoms that reside in the layers 2 and 3 (which would be represented by $\boldsymbol{r}_{2,3} \rightarrow - \boldsymbol{r}_{3,2}$). However, the crystallographic symmetry is not preserved if the AFM ordering of the magnetic moments in the Mn sites is taken into account because the magnetic moments that exist in the planes 1-4 and 2-3 are antiparallel with respect to each other within the exchange approximation. It is precisely this broken inversion symmetry that gives rise to the staggered SO field, $\boldsymbol{H}^{\mathrm{SO}}_i$, included in Eq. \eqref{eq:1}, in each type of magnetic sublattice, which allows to induce the AFM dynamics.

In this line, taking into account that Mn$_2$Au is a magnetically-based centro-asymmetric AFM, it is possible to introduce four vectors in the system, one FM vector, $\boldsymbol{m}_{\mathrm{a}}$, and three AFM vectors, $\boldsymbol{l}_i$, as linear combinations of the four sublattice magnetization vectors, $\boldsymbol{m}_i$, giving this as a result: $\boldsymbol{m}_{\mathrm{a}}=\left( \boldsymbol{m}_1+\boldsymbol{m}_2+\boldsymbol{m}_3+\boldsymbol{m}_4 \right)/4$, $\boldsymbol{l}_{\mathrm{a}}=\left( \boldsymbol{m}_1-\right.$ $\left. \boldsymbol{m}_2-\boldsymbol{m}_3+\boldsymbol{m}_4 \right)/4$, $\boldsymbol{l}_{\mathrm{b}}=\left( \boldsymbol{m}_1-\boldsymbol{m}_2+\boldsymbol{m}_3-\boldsymbol{m}_4 \right)/4$, and $\boldsymbol{l}_{\mathrm{c}}=\left( \boldsymbol{m}_1+\boldsymbol{m}_2-\boldsymbol{m}_3-\boldsymbol{m}_4 \right)/4$ \cite{turov2010symmetry}. In fact, one of these AFM vectors, namely $\boldsymbol{l}_{\mathrm{b}}$, can be chosen as the main one to define the system due to the specific magnetic symmetry of the Mn$_2$Au unit cell. At this point, we have to remember that planes 1-3 and 2-4 are magnetically identical, the relative magnetization direction being parallel to each other. Due to this, we can introduce a two sublattice model made up of Mn-based layers 2 and 3 (of type B and A represented in Fig. \ref{im:1} (b), respectively) assuming that $\boldsymbol{m}_1=\boldsymbol{m}_3$ and $\boldsymbol{m}_2=\boldsymbol{m}_4$. This has a result that the AFM vectors are given now by $\boldsymbol{l}_{{\mathrm{a}}, \mathrm{c}}=0$, $\boldsymbol{l}_{\mathrm{b}}=\left( \boldsymbol{m}_3-\boldsymbol{m}_2 \right)/2$, while the FM vector is represented by $\boldsymbol{m}_{\mathrm{a}}=\left( \boldsymbol{m}_3+\boldsymbol{m}_2 \right)/2$. This allows defining for Mn$_2$Au the main AFM vector as $\boldsymbol{l}_{\alpha}=\boldsymbol{l}_{\mathrm{b}}=\left( \boldsymbol{m}_3-\boldsymbol{m}_2 \right)/2$ and the magnetization vector as $\boldsymbol{m}_{\alpha}=\boldsymbol{m}_{\mathrm{a}}=\left( \boldsymbol{m}_3+\boldsymbol{m}_2 \right)/2$ \cite{turov2010symmetry}. In this way, it is possible to describe the dynamics in the layered AFM Mn$_2$Au through a two magnetic sublattice formalism taking into account only the two FM embedded layers 2 and 3, thus excluding from consideration the layers 1 and 4 for this purpose. We introduce also a more general definition of said variables in terms of the two types of magnetic layers of the system, A and B, as it is shown in Fig. \ref{im:1} (b), with which we have that $\boldsymbol{l}=\left( \boldsymbol{m}_{\mathrm{A}}-\boldsymbol{m}_{\mathrm{B}} \right)/2$ and $\boldsymbol{m}=\left( \boldsymbol{m}_{\mathrm{A}}+\boldsymbol{m}_{\mathrm{B}} \right)/2$, respectively, which is consistent with the $\boldsymbol{l}_{\alpha}$ and $\boldsymbol{m}_{\alpha}$ characterization.

To address the analytical description of the system, it is necessary to evaluate how many nearest neighbours exchange-based bonds have an impact on the inhomogeneous DW transition. For this, as it can be seen in Fig. \ref{im:1} (c), which shows a top view of the conventional unit cell, we will focus on the number of relevant first nearest neighbours for a Mn atom of layer 2 in an arbitrary position $x_i$ along the $x$-{\it th} spatial direction, which is characterized by the unit magnetization vector $\boldsymbol{m}_2 \left( x_i \right)$. In this line, it is possible to observe that said atom has four intersublattice first neighbours on layer 1, at a distance $a_0/2$ along the $x$-{\it th} axis, which are shown through $\boldsymbol{m}_1 (x_{i \pm 1/2}) $, which are mediated by the interaction characterized by the AFM $\mathcal{J}_1$ parameter. Also, the aforementioned atomic position has two first intrasublattice neighbours along the $x$-{\it th} spatial direction, at a distance $a_0$, characterized by $\boldsymbol{m}_2 (x_{i \pm 1})$, interacting through the FM exchange $\mathcal{J}_3$ contribution. It should be noted that both, the first intrasublattice neighbours in layer 2 along the $y$-{\it th} axis and the only intersublattice first nearest neighbour, mediated by the exchange interaction $\mathcal{J}_2$, of layer 3 along the $z$-{\it th} axis, are not taken into account because they do not impose any type of exchange penalty to determine the static or dynamic DW configuration in each of the sublattices of the system. In Suppl. Note I it can be found a simulations-based discussion about the role of the AFM exchange interaction $\mathcal{J}_2$ in this regard. On the other hand, due to the relative order of magnitude of the uniaxial magnetic anisotropy constants compared to the tetragonal ones, being given by $K_{2 \perp}/K_{4 \perp}=351$ and $K_{2 \parallel}/K_{4 \parallel}=7$, the fourth-order anisotropy constants will be neglected in the main approximation from now on.

In this way, we readjust the energy given by Eq. \eqref{eq:1} for the case of a two sublattice-based description in terms of the orthogonal set of unit vectors defined by $\boldsymbol{l}$ and $\boldsymbol{m}$, which satisfy the conditions  $\boldsymbol{m}^2+\boldsymbol{l}^2=1$ and $\boldsymbol{m} \cdot \boldsymbol{l}=0$. We take into account that even though these variables are constructed in terms of the magnetic sublattice types A and B, as discussed above, they refer from now on to the two FM embedded layers 2 and 3, in accordance with the definition of $\boldsymbol{m}_{\alpha}$ and $\boldsymbol{l}_{\alpha}$, the latter being explicitly represented in Fig. \ref{im:1} (b). However, in our case we are interested in describing the dynamics of a single magnetic texture, which occurs entirely in a single sublattice of the system. Since the introduction of the AFM vectors $\boldsymbol{m}$ and $\boldsymbol{l}$ implicitly assumes that the motion of a single DW occurs across both magnetic sublattices, we must halve the resulting magnetic anisotropy and Zeeman-like energies. Therefore, the energy, $w$, can be rewritten, taking into account the construction of the exchange part of the energy exposed in Suppl. Note II within an effective version of the nonlinear $\sigma$-model framework \cite{zvezdin1979pis, bar1985dynamics, lifshitz1980statistical}, due to the non-inclusion of the AFM exchange interaction along the c-axis of the system encoded by $\mathcal{J}_2$, in the exchange limit \cite{papanicolaou1995antiferromagnetic, tveten2016intrinsic}, as follows
\begin{equation}
w=\frac{1}{2} \, A \, {\boldsymbol{m}}^2+\frac{1}{8} \, a \, {\left( \partial_x \boldsymbol{l} \right)}^2+w_a \left( \boldsymbol{l} \right)-2 \gamma \hbar \, \, \boldsymbol{l} \cdot \boldsymbol{H}^{\mathrm{SO}}, \label{eq:2}
\end{equation}
where we introduced the homogeneous AFM exchange parameter, $A=16 \abs{\mathcal{J}_1}$, and the inhomogeneous FM-like exchange constant, given by $a=8 a^2_0 \left( \mathcal{J}_3+\abs{\mathcal{J}_1}/2 \right)$. Here the term $w_a \left( \boldsymbol{l} \right)$ encapsulates the uniaxial anisotropic contributions of the system, $w_a \left( \boldsymbol{l} \right)=\abs{K_{2 \perp}} \, {\left( \boldsymbol{l} \cdot \boldsymbol{\hat{z}} \right)}^2-K_{2 \parallel} \, {\left( \boldsymbol{l} \cdot \boldsymbol{\hat{y}} \right)}^2$, and  $\partial_x$ expresses the variation of the order parameter along the $x$-{\it th} spatial direction. This last statement is because it has been chosen that the current is injected along $\boldsymbol{j} \parallel \boldsymbol{\hat{x}}$, so $\boldsymbol{H}^{\mathrm{SO}}=H^{\mathrm{SO}}_y \boldsymbol{\hat{y}}$, inducing the DW motion along the $x$-{\it th} spatial direction. Also, we have rewritten the Zeeman-like term of Eq. \eqref{eq:1} taking into account that $\mu_0 \mu_s = 2 \gamma \hbar$, where $\gamma$ represents the gyromagnetic ratio and $\hbar$ is the reduced Planck constant.

Along the same line, it is possible to introduce how the Landau-Lifshitz-Gilbert (LLG) equations of the magnetic sublattice magnetization motions look like in the terms of  $\boldsymbol{l}$ and $\boldsymbol{m}$ vectors within the exchange limit \cite{kosevich1990magnetic, tveten2016intrinsic}, which are given by
\begin{gather}
\dot{\boldsymbol{l}}=\gamma \, \boldsymbol{H}^{\mathrm{eff}}_{\boldsymbol{m}} \times \boldsymbol{l}, \label{eq:3} \\
\dot{\boldsymbol{m}}=\left( \gamma \, \boldsymbol{H}^{\mathrm{eff}}_{\boldsymbol{l}}-\alpha \, \dot{\boldsymbol{l}} \right) \times \boldsymbol{l},
\label{eq:4}
\end{gather}
where $\boldsymbol{H}^{\mathrm{eff}}_{\boldsymbol{l}, \boldsymbol{m}}$ refers to the effective magnetic fields associated with the vector variables $\boldsymbol{l}, \boldsymbol{m}$. These effective fields can be expressed as $\boldsymbol{H}^{\mathrm{eff}}_{\boldsymbol{l}, \boldsymbol{m}}=- \frac{1}{2 \gamma \hbar} \, \frac{\delta w}{\delta \left( \boldsymbol{l}, \boldsymbol{m} \right)}$, where $\delta$ represents the functional derivative, $\alpha$ the phenomenological Gilbert damping parameter, which accounts for the dissipation processes, and the dot over a variable points out its derivative with respect to time, $t$. Through Eq. \eqref{eq:3}, it can be found that $\boldsymbol{m}=2 \hbar \, \left( \dot{\boldsymbol{l}} \times \boldsymbol{l} \right) /A$, expression which can be substituted in Eq. \eqref{eq:4} to obtain a second-order differential equation only in terms of the unit staggered AFM vector $\boldsymbol{l}$, which will be expressed by
\begin{equation}
\boldsymbol{l} \times \left[ \left( \partial^2_x \, \boldsymbol{l} \right)-\frac{1}{v^2_{\mathrm{m}}} \, \ddot{\boldsymbol{l}}+\boldsymbol{h}-\frac{4}{a} \, \frac{\partial  w_a \left( \boldsymbol{l} \right)}{\partial \boldsymbol{l}}-\eta \, \dot{\boldsymbol{l}} \right]=0,    
\label{eq:5}    
\end{equation}
where $v_{\mathrm{m}}$ represents the maximum magnon group velocity of the medium, which is given by $v_{\mathrm{m}}=\sqrt{a A}/ \left( 4 \hbar \right)=2 a_0 \sqrt{ 2 \abs{\mathcal{J}_1} \left( \mathcal{J}_3+\abs{\mathcal{J}_1}/2 \right)}/ \hbar=43.39$ km/s, $\boldsymbol{h}$ encodes the reduced SO field as $\boldsymbol{h}=8 \gamma \hbar \, \boldsymbol{H}^{\mathrm{SO}}/a$, and $\eta$ denotes the dissipative parameter expressed as $\eta=8\alpha \hbar/a$. Interestingly, it has been predicted that an AFM exchange interaction like $\mathcal{J}_2$, perpendicular to the inhomogeneous DW transition, should govern the value of $v_{\mathrm{m}}$ related to the low-frequency acoustic branch, contrary to what is demonstrated in Suppl. Notes I, both within and outside of the standard nonlinear $\sigma$-model framework \cite{zvezdin1999nonlinear,arana2017observation,yuan2018classification}.

In accordance with what it is shown in Fig. \ref{im:1} (d), it is possible to parameterize through spherical coordinates the unit N\'eel order parameter taking into account which is the in-plane easy-axis direction, giving rise to $\boldsymbol{l}=\left( \sin \varphi \, \cos \varepsilon , \cos \varphi \, \cos \varepsilon , - \sin \varepsilon \right)$, where $\varphi$ represents the azimuthal angle, which accounts for the rotation of the magnetization in the $xy$ plane being measured from the $y$-{\it th} axis, while $\varepsilon$ expresses the polar angle, which describes the out-of-plane canting being characterized from the $xy$ plane. Because we are working on the exchange limit, it is possible to assume that $\varepsilon \simeq 0$, whereby the reduced AFM vector can be expressed as $\boldsymbol{l} \simeq \left( \sin \varphi , \cos \varphi, 0 \right)$. This being the case, it is possible to reduce Eq. \eqref{eq:5} to a sine-Gordon wave-like equation \cite{bar1980nonlinear, andreev1980symmetry}, with the following functional form
\begin{equation}
\frac{1}{v^2_{\mathrm{m}}} \, \ddot{\varphi}-\left( \partial^2_x \varphi \right)+\frac{1}{2 \Delta^2_0} \, \sin 2 \varphi + h \, \sin \varphi=-\eta \, \dot{\varphi},
\label{eq:6}    
\end{equation}
where $\Delta_0$ stands for the DW width at rest, which is given by $\Delta_0 =\sqrt{a / \left( 8 K_{2 \parallel} \right)} = a_0 \sqrt{\left( \mathcal{J}_3+\abs{\mathcal{J}_1}/2 \right)/ K_{2 \parallel}} = 19.17$ nm, and where $h=8 \gamma \hbar \, H^{\mathrm{SO}}_y/ \, a$ denotes the reduced scalar SO field.

At this point, it is convenient to consider the magnetic texture dynamics within the framework of the well-known collective coordinates approach \cite{rajaraman1982solitons}. For this, it is usual to introduce what is known as Walker-like rigid profile through the angular variable that defines the spatio-temporal evolution of the magnetization, $\varphi \left( x, t \right)=2 \, \arctan \, \mathrm{exp} \left[ \left( x- q \left( t \right) \right) /\Delta \left( t \right) \right]$ \cite{schryer1974motion}, with $q$ being the DW center position and $\Delta$ being the dynamic DW width. Due to the Lorentz invariance shown by Eqs. \eqref{eq:5} and \eqref{eq:6}, the magnetic soliton dynamics in AFM shows emergent special relativity signatures. In particular, the DW width decreases as the velocity of the magnetic texture, $\dot{q}$, increases, which is given by the expression $\Delta \left( t \right) = \Delta_0 \, \beta \left( t \right)$, where $\beta \left( t \right) = \sqrt{1-{\left( \dot{q} \left( t \right) / v_{\mathrm{m}} \right)}^2}$ represents the Lorentz factor. In order to avoid the excitation of internal modes of the magnetic texture, we focus on quasistatic processes, being its spatial extent variation governed entirely by the special relativity-based Lorentz factor, so we can neglect the time derivatives of the DW width $\Delta$. In this way, we obtain that
\begin{equation}
\frac{1}{\Delta \, v^2_{\mathrm{m}}} \, \ddot{q}+\frac{\eta}{\Delta} \, \dot{q}-h=0.
\label{eq:7} 
\end{equation}

As it can be seen in Eq. \eqref{eq:7}, we have a Newton-like second-order differential equation for the time evolution of the DW center position, $q$, which explicitly shows the inertial nature of the magnetic texture \cite{tveten2013staggered, yuan2018classification}. In the particular case in which a constant SO field is applied, it is  possible to access a steady-state-like DW motion regime after the accommodation of the soliton to its new dynamic state. In this sense, we can reduce the previous equation to a compact expression that accounts for the steady-state DW velocity, which we denote from now on as $v=\dot{q}$, which is given by
\begin{equation}
v= \frac{v_{\mathrm{m}}}{\sqrt{1+{\left( v_{\mathrm{m}}/ v_0 \right)}^2}} \, ,
\label{eq:8}
\end{equation}
where $v_0=h \Delta_0/\eta$. 

\section{Relativistic and inertial domain wall dynamic signatures}\label{section:comparison}

To verify the predictions obtained above through our effective version of the nonlinear $\sigma$-model, we have performed atomistic spin dynamics simulations of the real crystallographic and magnetic conventional unit cell of Mn$_2$Au, as it is shown in Fig. \ref{im:1} (a), taking into account all interactions of the system, as it is collected in Eq. \eqref{eq:1}. With this objective in mind, the system of the coupled LLG equations of motion of the local magnetic moments $ \boldsymbol{m}_i$, being given by

\begin{figure}[!ht]
\includegraphics[width=8cm]{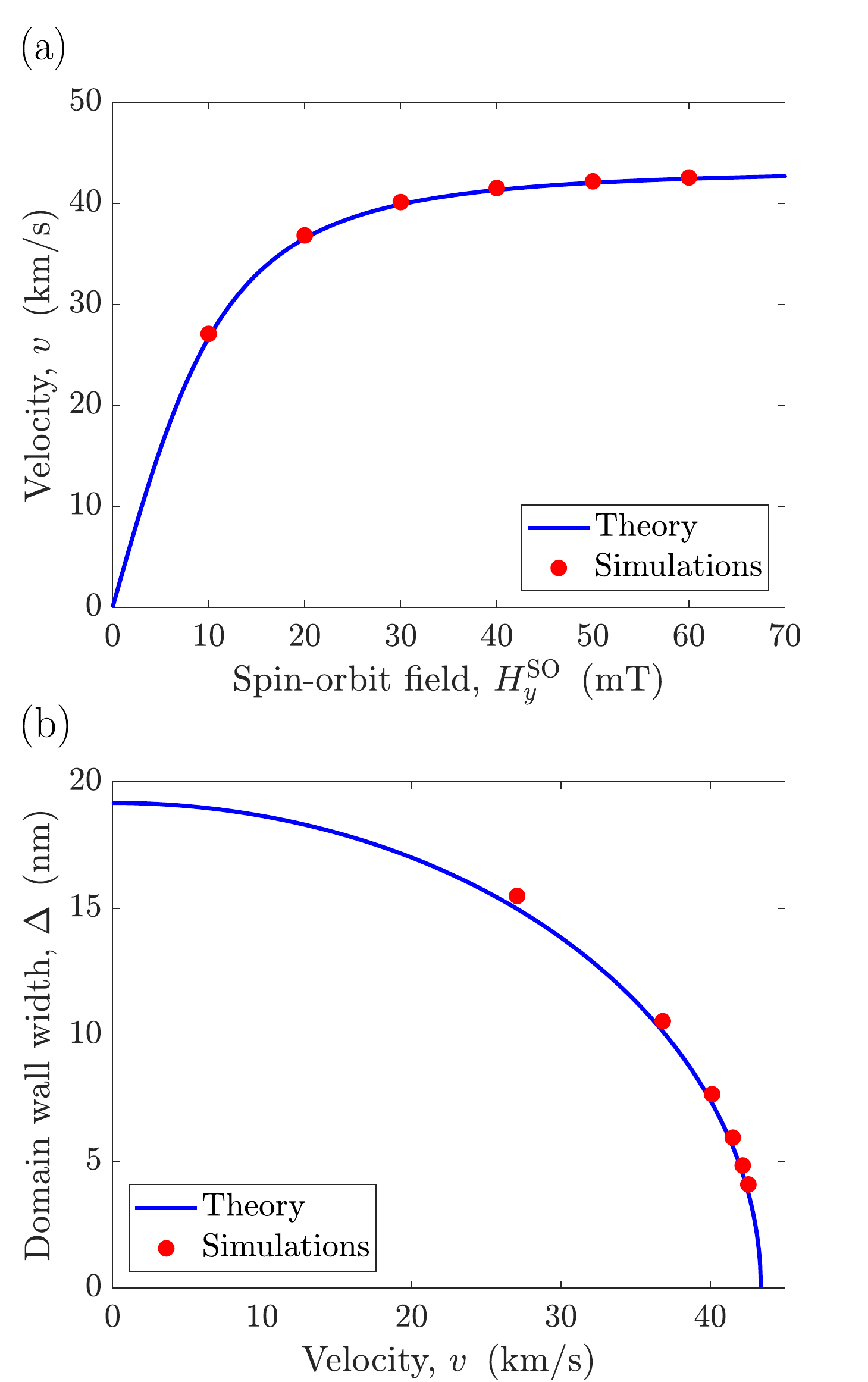}
\caption{Comparison of the relativistic signatures for steady-state processes in Mn$_2$Au obtained through atomistic spin dynamics simulations and theory. (a) Saturation of the velocity, $v$, of the magnetic texture as the SO field, $H^{\mathrm{SO}}_y$, increases, being based the analytical formalism in Eq. \eqref{eq:8}. (b) DW width, $\Delta$, contraction as the speed $v$ of the magnetic soliton increases, the theoretical prediction coming from the combination of the relativistic expression $\Delta= \Delta_0 \, \beta$ and Eq. \eqref{eq:8}.} 
\label{im:2}
\end{figure}
\begin{figure*}[hbt]
\centering
\includegraphics[width=18cm]{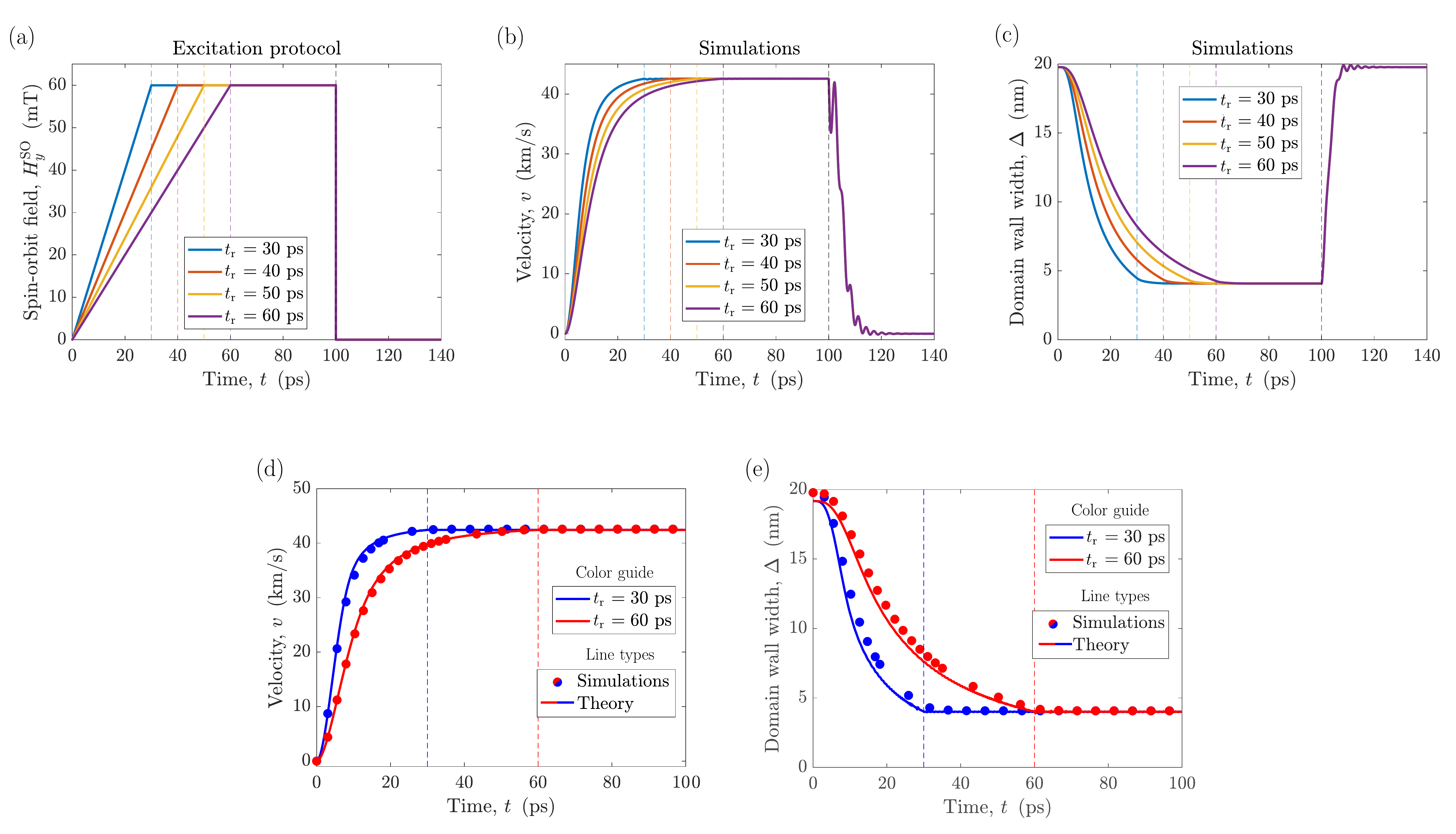}
\caption{(a) Time-dependent staggered SO field-based excitation protocol, $H^{\mathrm{SO}}_ y \left( t \right)$, applied in each Mn-based FM layer of Mn$_2$Au for different ramping times, $t_{\mathrm{r}}$. Dynamic time evolution of the DW velocity $v$ (b) and width $\Delta$ (c) obtained through atomistic spin dynamics simulations for different ramping times $t_{\mathrm{r}}$. Comparison of the dynamic time evolution of the velocity $v$ (d) and width $\Delta$ (e) of a N\'eel-like DW for atomistic spin dynamics simulations and the analytical expressions given by Eq. \eqref{eq:7} and $\Delta=\Delta_0 \, \beta$ for two ramping times, $t_{\mathrm{r}}=30$ and $60$ ps. Each vertical colored dashed line represents the end of the ramped process for the different ramping times $t_{\mathrm{r}}$, while the dashed black line denotes the instant $t=100$ ps at which the driving SO field $H^{\mathrm{SO}}_y$ is abruptly turned off.}
\label{im:3}
\end{figure*}
\begin{equation}
\frac{\left( 1+\alpha^2 \right)}{\gamma} \, \dot{\boldsymbol{m}}_i=- \boldsymbol{m}_i \times \boldsymbol{H}^{\mathrm{eff}}_i-\alpha \, \boldsymbol{m}_i \times \left( \boldsymbol{m}_i \times \boldsymbol{H}^{\mathrm{eff}}_i \right),
\label{eq:9}
\end{equation}
is evaluated numerically site by site through a fifth-order Runge-Kutta method. Here, $\boldsymbol{H}^{\mathrm{eff}}_i$ represents the effective field at each lattice position, which depends on the interactions exposed in Eq. \eqref{eq:1}, as $\boldsymbol{H}^{\mathrm{eff}}_i=-\frac{1}{\mu_0 \mu_{\mathrm{s}}} \frac{\delta w}{\delta \boldsymbol{m}_i}$, and the damping parameter is $\alpha=0.001$ \cite{gomonay2018narrow,otxoa2020giant,otxoa2020walker}. In this case, the computational domain consists of 60000 cells along the $x$-{\it th} propagation direction, one cell width with periodic boundary conditions along $y$-{\it th} direction, and one cell thick along the $z$-{\it th} direction \cite{clarification}. Due to their simple functional forms in terms of intrinsic parameters of the material, we can test the validity of our analytical formalism by comparing the values of the DW width at rest, $\Delta_0$, and the maximum magnon group velocity of the medium, $v_{\mathrm{m}}$, with the simulated ones. This being the case, we have found that the simulated values are $\Delta_0=19.78$ nm and $v_{\mathrm{m}}=43.3$ km/s. The theoretically-predicted DW width at rest $\Delta_0$ presents a good correspondence with the fitted rigid DW profile-based simulated value, differing only by a $3.1 \%$, which possibly comes from the non-inclusion in the analytical model, for simplicity, of the in-plane tetragonal anisotropy contribution encoded by $K_{4 \parallel}$. On the other hand, the maximum magnon group velocity $v_{\mathrm{m}}$ obtained analytically and by simulations coincide in a $99.93 \%$. With this satisfactory correspondence between the theory and simulations, which supports the non-inclusion in our formalism of the AFM exchange interaction given by $\mathcal{J}_2$, we can explore the emergent special relativity signatures during steady-state DW dynamic processes. As it can be seen in Fig. \ref{im:2}, the saturation of the magnetic texture velocity, $v$, as it is predicted by Eq. \eqref{eq:8}, and the contraction of the DW width, $\Delta$, in correspondence with the expression $\Delta= \Delta_0 \, \beta$, as the SO field, $H^{\mathrm{SO}}_y$, increases are verified.

To explore the inertial signatures on our system, we use the time-dependent SO field-based excitation regimes represented in Fig. \ref{im:3} (a). As it can be seen, there are three well differentiated regions. In the first one, which covers the interval $t \in \left[ \left. 0, t_{\mathrm{r}} \right) \right.$, being $t_{\mathrm{r}}$ the time taken to reach a constant value of the field of $H^{\mathrm{SO}}_y=60$ mT, which we denote as ramping time, the rest state of the magnetic texture is disturbed through a SO field that increases linearly with time. We denote this regime as region I, and each colored dashed line in Fig. \ref{im:3} corresponds to a certain $ t_{\mathrm{r}} $ that defines the aforementioned domain. Consistently with Eq. \eqref{eq:7}, which implicitly shows the existence of a non-zero DW mass, the initial response of the magnetic texture to the external stimulus is fast, but not instantaneous, as it can be seen in Figs. \ref{im:3} (b, c). At the time when a constant field value of $60$ mT is reached, that is, at $ t = t_{\mathrm{r}} $, the magnetic soliton tends to a steady-state regime (region II), which covers the interval $t \in \left[ \left. t_{\mathrm{r}}, 100 \, \, \mathrm{ps} \right) \right.$. This upper limit is denoted by a black dashed line when appropriate in Fig. \ref{im:3}. Thus, it is possible to observe in Figs. \ref{im:3} (b, c) that, in the region II, after a brief adaptation period to the new dynamic regime, which is a sample of the inertial nature of the process, the magnetic texture moves steadily at a speed of $v=42.56$ km/s. This is very close to the maximum magnon group velocity of the medium, denoting a $98\%$ of it, while shrinking to a width of $\Delta=4.08$ nm, which represents a contraction of $80\%$ with respect to the simulated DW width at rest. Finally, at a certain moment given by $t=100$ ps, the SO field is abruptly switched off. This makes it possible to observe that the magnetic texture is capable of initiating an after-pulse displacement in the absence of an external stimulus at the same time as its width expands until it stops completely, as it can be seen in Figs. \ref{im:3} (b, c). We denote this regime as region III, and it covers the interval $t \in \left[ 100, 140 \right]$ ps.

\begin{figure}[!ht]
\includegraphics[width=8cm]{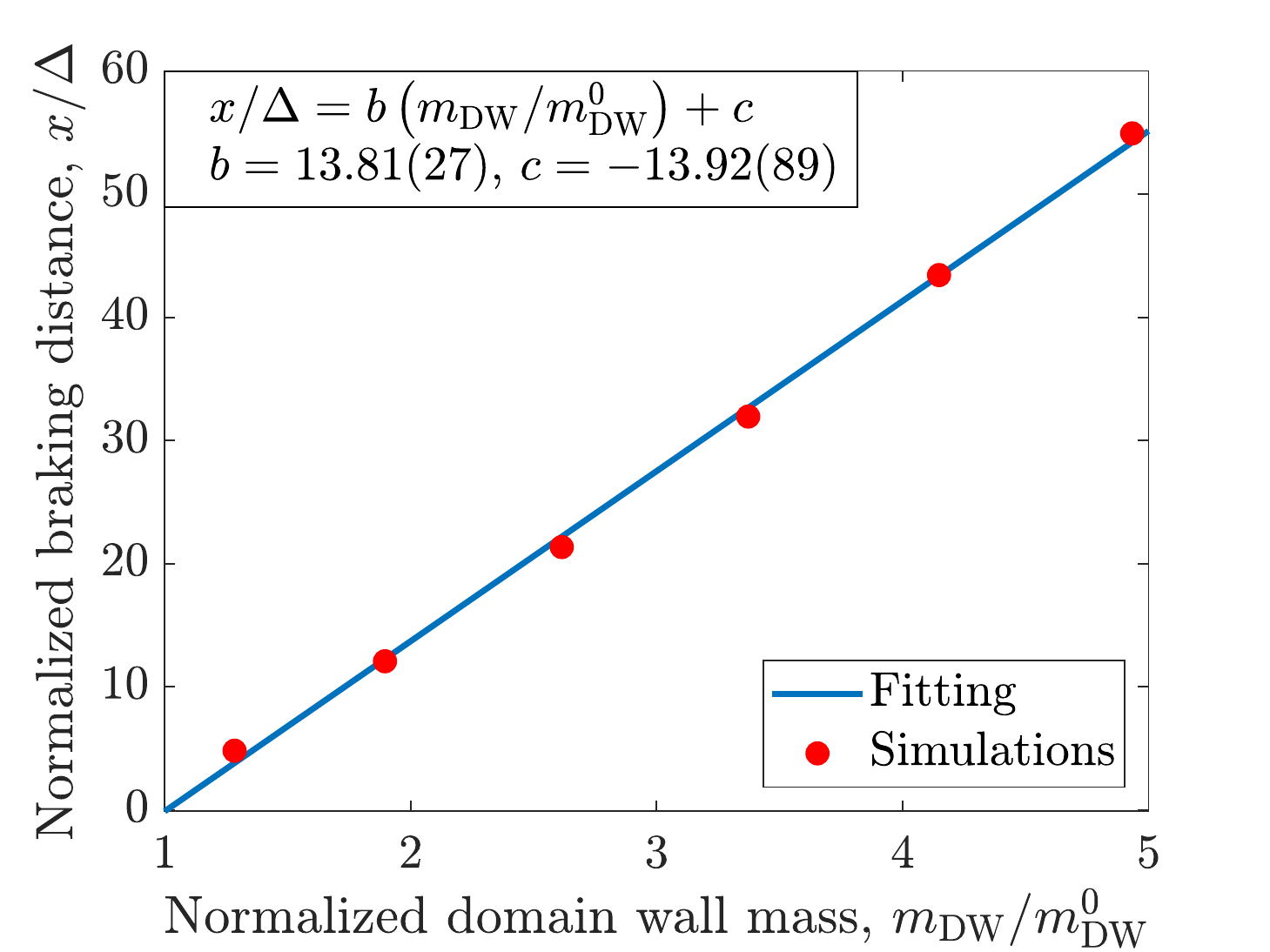}
\caption{Quasilinear correspondence between the normalized relativistic DW mass, $m_{\mathrm{DW}}/m^0_{\mathrm{DW}}$, where $m_{\mathrm{DW}}$ and $m^0_{\mathrm{DW}}$ represent the aforementioned mass in steady-state and at rest, respectively, and the normalized braking distance, $x$, travelled by the magnetic texture once the SO field is turned off abruptly, $x/\Delta$, in terms of the steady-state DW width, $\Delta$, characterized by a linear fitting where $b$ and $c$ represent the adjustment parameters, together with their corresponding uncertainties.}
\label{im:4}
\end{figure}

The range of values considered for the ramping time $t_{\mathrm{r}}$ has been chosen to avoid the excitation of internal DW modes during the acceleration process, in such a way that the simulations were comparable to the scenario exposed in Section \ref{section:theory} through Eq. \eqref{eq:7}. As it can be seen in Figs. \ref{im:3} (d, e), in these circumstances there is a great correspondence between the simulated and the theoretically-predicted velocity of the magnetic texture, $v$, and its spatial extent, $\Delta$, in regions I and II. However, if attention is paid to Figs. \ref{im:3} (b, c) in the region III, it is possible to appreciate undulations once the SO field $H^{\mathrm{SO}}_y$ is turned off. These ripples observed in the simulated DW velocity and width in the region III cause longer decay times than those predicted through a simple Newton-like pseudoparticle behaviour. Therefore, we avoid the analytical evaluation of this region through Eq. \eqref{eq:7} because the magnetic soliton is no longer fulfilling the imposed rigid profile constraint. However, the after-pulse displacement that the magnetic texture experiences in region III is related to the increase in the exchange-based relativistic DW mass obtained during the acceleration process in region I while its width shrinked. This is a purely inertial phenomenon, which is consistent with the massive pseudoparticle behavior captured by Eq. \eqref{eq:7}, in which the higher is the dynamic DW mass after turning off the external stimulus, the greater is the braking distance travelled by it. Interestingly, it is the fact that the magnetic soliton moves along a particular direction which sets how this stored relativistic exchange energy is transformed into a translational displacement, manifesting its massive particle-like behavior rather than being dissipated in a breather-like fashion or through the emission of spin waves without prolonging its mobility. For different values of the SO field during region II, we have found through simulations that there is a quasilinear relationship between the after-pulse distance travelled by the magnetic texture, $x$, normalized to the its steady-state DW width, $\Delta$, and its steady-state DW mass, $m_{\mathrm{DW}}$, normalized to its rest state value, $m^0_{\mathrm{DW}}$, this is, $m_{\mathrm{DW}}/m^0_{\mathrm{DW}}=1/\beta$ according to Eq. \eqref{eq:7}, which can be seen in Fig. \ref{im:4} and can be expressed as
\begin{equation}
\frac{x}{\Delta}=b \, \, \frac{m_{\mathrm{DW}}}{m^0_{\mathrm{DW}}}+c,
\label{eq:10}
\end{equation}
where $b$ and $c$ are two fitting-dependent parameters, which are given, accompanied by their associated uncertainties, by $b=13.81(27)$ and $c=-13.92(89)$. It is remarkable that the DW is capable of undertaking exchange-based after-pulse displacements of the order of $4$ to $11$ times greater than the DW width at rest, $\Delta_0$, for SO fields between $10$-$60$ mT. This accurate prediction of the braking distance experienced by the magnetic soliton opens the door to implement low-energy consumption processes.

\section{Conclusions}\label{section:conclusions}

We addressed the theoretical characterization of the dynamic evolution of a 1D N\'eel-like DW in one of the FM sublattices of the layered collinear AFM Mn$_2$Au driven by the current-induced SO fields. Despite the complexity of the system, we have exploited the symmetric inequivalence between the crystallographic and the magnetic unit cell to reduce its description to a two-sublattice problem. Since the AFM exchange interaction directed along the c-axis of the system, which is encoded through $\mathcal{J}_2$, has a null projection along the 1D inhomogeneous magnetic texture transition, it has no impact on the temperature-independent standard nonlinear $\sigma$-model. Because of this, we have worked on within an effective theory framework avoiding its inclusion, methodology which can be extrapolated to layered multisublattice AFM with different exchange-oriented contributions. In the rigid profile approximation, we have shown that it is possible to reduce the dynamic description to a Newton-like second-order differential equation of motion. Comparing our formalism with atomistic spin dynamics simulations, we have been able to replicate with a high degree of precision the relativistic and inertial signatures of the magnetic texture motion during quasistatic dynamic processes within the framework of our effective model. After the abrupt shutdown of the SO field in simulations, the rigid DW profile approach is no longer supported and our analytical formalism fails to describe the after-pulse inertial dynamic regime. Interestingly, during the deceleration process the relativistic exchange energy accumulated during the previous dynamic evolution of the magnetic texture is converted into translational mobility, rather than being released in a breather-like fashion or through the emission of spin waves with non-associated displacement. We have found a quasilinear relationship that allows us to predict, for the range of simulated SO fields, the value of the braking distance travelled by the DW through the knowledge of its relativistic mass before turning off the external stimulus. This detailed dynamic characterization of the 1D magnetic texture in the complex multisublattice antiferromagnet Mn$_2$Au is of potential interest for AFM DW-based technology applications.

\section*{ACKNOWLEDGEMENTS}\label{section:acknowledgements}

R.R.-E., K.Y.G., and R.M.O. thanks O. Chubykalo-Fesenko, S. Khmelevskyi, A. A. Sapozhnik, M. Jourdan, A. K. Zvezdin, and B. A. Ivanov for the fruitful discussions that have helped us to improve this manuscript. The work of R.M.O. and K.Y.G. was partially supported by the STSM Grants from the COST Action CA17123 ``Ultrafast opto-magneto-electronics for non-dissipative information technology". K.Y.G. acknowledges support by IKERBASQUE (the Basque Foundation for Science) and the Spanish Ministry of Science and Innovation under grant PID2019-108075RB-C33/AEI/10.13039/501100011033.

\bibliographystyle{apsrev4-1}
\bibliography{Bibliography}

\begin{thebibliography}{44}%
\makeatletter
\providecommand \@ifxundefined [1]{%
 \@ifx{#1\undefined}
}%
\providecommand \@ifnum [1]{%
 \ifnum #1\expandafter \@firstoftwo
 \else \expandafter \@secondoftwo
 \fi
}%
\providecommand \@ifx [1]{%
 \ifx #1\expandafter \@firstoftwo
 \else \expandafter \@secondoftwo
 \fi
}%
\providecommand \natexlab [1]{#1}%
\providecommand \enquote  [1]{``#1''}%
\providecommand \bibnamefont  [1]{#1}%
\providecommand \bibfnamefont [1]{#1}%
\providecommand \citenamefont [1]{#1}%
\providecommand \href@noop [0]{\@secondoftwo}%
\providecommand \href [0]{\begingroup \@sanitize@url \@href}%
\providecommand \@href[1]{\@@startlink{#1}\@@href}%
\providecommand \@@href[1]{\endgroup#1\@@endlink}%
\providecommand \@sanitize@url [0]{\catcode `\\12\catcode `\$12\catcode
  `\&12\catcode `\#12\catcode `\^12\catcode `\_12\catcode `\%12\relax}%
\providecommand \@@startlink[1]{}%
\providecommand \@@endlink[0]{}%
\providecommand \url  [0]{\begingroup\@sanitize@url \@url }%
\providecommand \@url [1]{\endgroup\@href {#1}{\urlprefix }}%
\providecommand \urlprefix  [0]{URL }%
\providecommand \Eprint [0]{\href }%
\providecommand \doibase [0]{http://dx.doi.org/}%
\providecommand \selectlanguage [0]{\@gobble}%
\providecommand \bibinfo  [0]{\@secondoftwo}%
\providecommand \bibfield  [0]{\@secondoftwo}%
\providecommand \translation [1]{[#1]}%
\providecommand \BibitemOpen [0]{}%
\providecommand \bibitemStop [0]{}%
\providecommand \bibitemNoStop [0]{.\EOS\space}%
\providecommand \EOS [0]{\spacefactor3000\relax}%
\providecommand \BibitemShut  [1]{\csname bibitem#1\endcsname}%
\let\auto@bib@innerbib\@empty
\bibitem [{\citenamefont {Gomonay}\ \emph {et~al.}(2016)\citenamefont
  {Gomonay}, \citenamefont {Jungwirth},\ and\ \citenamefont
  {Sinova}}]{gomonay2016high}%
  \BibitemOpen
  \bibfield  {author} {\bibinfo {author} {\bibfnamefont {O.}~\bibnamefont
  {Gomonay}}, \bibinfo {author} {\bibfnamefont {T.}~\bibnamefont {Jungwirth}},
  \ and\ \bibinfo {author} {\bibfnamefont {J.}~\bibnamefont {Sinova}},\ }\href
  {https://journals.aps.org/prl/abstract/10.1103/PhysRevLett.117.017202}
  {\bibfield  {journal} {\bibinfo  {journal} {Phys. Rev. Lett.}\ }\textbf
  {\bibinfo {volume} {117}},\ \bibinfo {pages} {017202} (\bibinfo {year}
  {2016})}\BibitemShut {NoStop}%
\bibitem [{\citenamefont {Shiino}\ \emph {et~al.}(2016)\citenamefont {Shiino},
  \citenamefont {Oh}, \citenamefont {Haney}, \citenamefont {Lee}, \citenamefont
  {Go}, \citenamefont {Park},\ and\ \citenamefont
  {Lee}}]{shiino2016antiferromagnetic}%
  \BibitemOpen
  \bibfield  {author} {\bibinfo {author} {\bibfnamefont {T.}~\bibnamefont
  {Shiino}}, \bibinfo {author} {\bibfnamefont {S.-H.}\ \bibnamefont {Oh}},
  \bibinfo {author} {\bibfnamefont {P.~M.}\ \bibnamefont {Haney}}, \bibinfo
  {author} {\bibfnamefont {S.-W.}\ \bibnamefont {Lee}}, \bibinfo {author}
  {\bibfnamefont {G.}~\bibnamefont {Go}}, \bibinfo {author} {\bibfnamefont
  {B.-G.}\ \bibnamefont {Park}}, \ and\ \bibinfo {author} {\bibfnamefont
  {K.-J.}\ \bibnamefont {Lee}},\ }\href
  {https://journals.aps.org/prl/abstract/10.1103/PhysRevLett.117.087203}
  {\bibfield  {journal} {\bibinfo  {journal} {Phys. Rev. Lett.}\ }\textbf
  {\bibinfo {volume} {117}},\ \bibinfo {pages} {087203} (\bibinfo {year}
  {2016})}\BibitemShut {NoStop}%
\bibitem [{\citenamefont {Yang}\ \emph {et~al.}(2019)\citenamefont {Yang},
  \citenamefont {Yuan}, \citenamefont {Yan}, \citenamefont {Zhang},\ and\
  \citenamefont {Yan}}]{yang2019atomic}%
  \BibitemOpen
  \bibfield  {author} {\bibinfo {author} {\bibfnamefont {H.}~\bibnamefont
  {Yang}}, \bibinfo {author} {\bibfnamefont {H.~Y.}\ \bibnamefont {Yuan}},
  \bibinfo {author} {\bibfnamefont {M.}~\bibnamefont {Yan}}, \bibinfo {author}
  {\bibfnamefont {H.~W.}\ \bibnamefont {Zhang}}, \ and\ \bibinfo {author}
  {\bibfnamefont {P.}~\bibnamefont {Yan}},\ }\href
  {https://journals.aps.org/prb/abstract/10.1103/PhysRevB.100.024407}
  {\bibfield  {journal} {\bibinfo  {journal} {Phys. Rev. B}\ }\textbf {\bibinfo
  {volume} {100}},\ \bibinfo {pages} {024407} (\bibinfo {year}
  {2019})}\BibitemShut {NoStop}%
\bibitem [{\citenamefont {Otxoa}\ \emph
  {et~al.}(2020{\natexlab{a}})\citenamefont {Otxoa}, \citenamefont {Roy},
  \citenamefont {Rama-Eiroa}, \citenamefont {Godinho}, \citenamefont
  {Guslienko},\ and\ \citenamefont {Wunderlich}}]{otxoa2020walker}%
  \BibitemOpen
  \bibfield  {author} {\bibinfo {author} {\bibfnamefont {R.~M.}\ \bibnamefont
  {Otxoa}}, \bibinfo {author} {\bibfnamefont {P.~E.}\ \bibnamefont {Roy}},
  \bibinfo {author} {\bibfnamefont {R.}~\bibnamefont {Rama-Eiroa}}, \bibinfo
  {author} {\bibfnamefont {J.}~\bibnamefont {Godinho}}, \bibinfo {author}
  {\bibfnamefont {K.~Y.}\ \bibnamefont {Guslienko}}, \ and\ \bibinfo {author}
  {\bibfnamefont {J.}~\bibnamefont {Wunderlich}},\ }\href
  {https://www.nature.com/articles/s42005-020-00456-5} {\bibfield  {journal}
  {\bibinfo  {journal} {Commun. Phys.}\ }\textbf {\bibinfo {volume} {3}},\
  \bibinfo {pages} {190} (\bibinfo {year} {2020}{\natexlab{a}})}\BibitemShut
  {NoStop}%
\bibitem [{\citenamefont {Kim}\ \emph {et~al.}(2014)\citenamefont {Kim},
  \citenamefont {Tserkovnyak},\ and\ \citenamefont
  {Tchernyshyov}}]{kim2014propulsion}%
  \BibitemOpen
  \bibfield  {author} {\bibinfo {author} {\bibfnamefont {S.~K.}\ \bibnamefont
  {Kim}}, \bibinfo {author} {\bibfnamefont {Y.}~\bibnamefont {Tserkovnyak}}, \
  and\ \bibinfo {author} {\bibfnamefont {O.}~\bibnamefont {Tchernyshyov}},\
  }\href {https://journals.aps.org/prb/abstract/10.1103/PhysRevB.90.104406}
  {\bibfield  {journal} {\bibinfo  {journal} {Phys. Rev. B}\ }\textbf {\bibinfo
  {volume} {90}},\ \bibinfo {pages} {104406} (\bibinfo {year}
  {2014})}\BibitemShut {NoStop}%
\bibitem [{\citenamefont {Tatara}\ \emph {et~al.}(2020)\citenamefont {Tatara},
  \citenamefont {Akosa},\ and\ \citenamefont {Otxoa}}]{tatara2020magnon}%
  \BibitemOpen
  \bibfield  {author} {\bibinfo {author} {\bibfnamefont {G.}~\bibnamefont
  {Tatara}}, \bibinfo {author} {\bibfnamefont {C.~A.}\ \bibnamefont {Akosa}}, \
  and\ \bibinfo {author} {\bibfnamefont {R.~M.}\ \bibnamefont {Otxoa}},\ }\href
  {https://journals.aps.org/prresearch/abstract/10.1103/PhysRevResearch.2.043226}
  {\bibfield  {journal} {\bibinfo  {journal} {Phys. Rev. Res.}\ }\textbf
  {\bibinfo {volume} {2}},\ \bibinfo {pages} {043226} (\bibinfo {year}
  {2020})}\BibitemShut {NoStop}%
\bibitem [{\citenamefont {Tveten}\ \emph {et~al.}(2013)\citenamefont {Tveten},
  \citenamefont {Qaiumzadeh}, \citenamefont {Tretiakov},\ and\ \citenamefont
  {Brataas}}]{tveten2013staggered}%
  \BibitemOpen
  \bibfield  {author} {\bibinfo {author} {\bibfnamefont {E.~G.}\ \bibnamefont
  {Tveten}}, \bibinfo {author} {\bibfnamefont {A.}~\bibnamefont {Qaiumzadeh}},
  \bibinfo {author} {\bibfnamefont {O.~A.}\ \bibnamefont {Tretiakov}}, \ and\
  \bibinfo {author} {\bibfnamefont {A.}~\bibnamefont {Brataas}},\ }\href
  {https://journals.aps.org/prl/abstract/10.1103/PhysRevLett.110.127208}
  {\bibfield  {journal} {\bibinfo  {journal} {Phys. Rev. Lett.}\ }\textbf
  {\bibinfo {volume} {110}},\ \bibinfo {pages} {127208} (\bibinfo {year}
  {2013})}\BibitemShut {NoStop}%
\bibitem [{\citenamefont {Yuan}\ \emph {et~al.}(2018)\citenamefont {Yuan},
  \citenamefont {Wang}, \citenamefont {Yung},\ and\ \citenamefont
  {Wang}}]{yuan2018classification}%
  \BibitemOpen
  \bibfield  {author} {\bibinfo {author} {\bibfnamefont {H.~Y.}\ \bibnamefont
  {Yuan}}, \bibinfo {author} {\bibfnamefont {W.}~\bibnamefont {Wang}}, \bibinfo
  {author} {\bibfnamefont {M.-H.}\ \bibnamefont {Yung}}, \ and\ \bibinfo
  {author} {\bibfnamefont {X.~R.}\ \bibnamefont {Wang}},\ }\href
  {https://journals.aps.org/prb/abstract/10.1103/PhysRevB.97.214434} {\bibfield
   {journal} {\bibinfo  {journal} {Phys. Rev. B}\ }\textbf {\bibinfo {volume}
  {97}},\ \bibinfo {pages} {214434} (\bibinfo {year} {2018})}\BibitemShut
  {NoStop}%
\bibitem [{\citenamefont {Selzer}\ \emph {et~al.}(2016)\citenamefont {Selzer},
  \citenamefont {Atxitia}, \citenamefont {Ritzmann}, \citenamefont {Hinzke},\
  and\ \citenamefont {Nowak}}]{selzer2016inertia}%
  \BibitemOpen
  \bibfield  {author} {\bibinfo {author} {\bibfnamefont {S.}~\bibnamefont
  {Selzer}}, \bibinfo {author} {\bibfnamefont {U.}~\bibnamefont {Atxitia}},
  \bibinfo {author} {\bibfnamefont {U.}~\bibnamefont {Ritzmann}}, \bibinfo
  {author} {\bibfnamefont {D.}~\bibnamefont {Hinzke}}, \ and\ \bibinfo {author}
  {\bibfnamefont {U.}~\bibnamefont {Nowak}},\ }\href
  {https://journals.aps.org/prl/abstract/10.1103/PhysRevLett.117.107201}
  {\bibfield  {journal} {\bibinfo  {journal} {Phys. Rev. Lett.}\ }\textbf
  {\bibinfo {volume} {117}},\ \bibinfo {pages} {107201} (\bibinfo {year}
  {2016})}\BibitemShut {NoStop}%
\bibitem [{\citenamefont {{\v{Z}}elezn{\`y}}\ \emph {et~al.}(2014)\citenamefont
  {{\v{Z}}elezn{\`y}}, \citenamefont {Gao}, \citenamefont {V{\`y}born{\`y}},
  \citenamefont {Zemen}, \citenamefont {Ma{\v{s}}ek}, \citenamefont {Manchon},
  \citenamefont {Wunderlich}, \citenamefont {Sinova},\ and\ \citenamefont
  {Jungwirth}}]{vzelezny2014relativistic}%
  \BibitemOpen
  \bibfield  {author} {\bibinfo {author} {\bibfnamefont {J.}~\bibnamefont
  {{\v{Z}}elezn{\`y}}}, \bibinfo {author} {\bibfnamefont {H.}~\bibnamefont
  {Gao}}, \bibinfo {author} {\bibfnamefont {K.}~\bibnamefont
  {V{\`y}born{\`y}}}, \bibinfo {author} {\bibfnamefont {J.}~\bibnamefont
  {Zemen}}, \bibinfo {author} {\bibfnamefont {J.}~\bibnamefont {Ma{\v{s}}ek}},
  \bibinfo {author} {\bibfnamefont {A.}~\bibnamefont {Manchon}}, \bibinfo
  {author} {\bibfnamefont {J.}~\bibnamefont {Wunderlich}}, \bibinfo {author}
  {\bibfnamefont {J.}~\bibnamefont {Sinova}}, \ and\ \bibinfo {author}
  {\bibfnamefont {T.}~\bibnamefont {Jungwirth}},\ }\href
  {https://journals.aps.org/prl/abstract/10.1103/PhysRevLett.113.157201}
  {\bibfield  {journal} {\bibinfo  {journal} {Phys. Rev. Lett.}\ }\textbf
  {\bibinfo {volume} {113}},\ \bibinfo {pages} {157201} (\bibinfo {year}
  {2014})}\BibitemShut {NoStop}%
\bibitem [{\citenamefont {Wadley}\ \emph {et~al.}(2016)\citenamefont {Wadley},
  \citenamefont {Howells}, \citenamefont {{\v{Z}}elezn{\`y}}, \citenamefont
  {Andrews}, \citenamefont {Hills}, \citenamefont {Campion}, \citenamefont
  {Nov{\'a}k}, \citenamefont {Olejn{\'\i}k}, \citenamefont {Maccherozzi},
  \citenamefont {Dhesi} \emph {et~al.}}]{wadley2016electrical}%
  \BibitemOpen
  \bibfield  {author} {\bibinfo {author} {\bibfnamefont {P.}~\bibnamefont
  {Wadley}}, \bibinfo {author} {\bibfnamefont {B.}~\bibnamefont {Howells}},
  \bibinfo {author} {\bibfnamefont {J.}~\bibnamefont {{\v{Z}}elezn{\`y}}},
  \bibinfo {author} {\bibfnamefont {C.}~\bibnamefont {Andrews}}, \bibinfo
  {author} {\bibfnamefont {V.}~\bibnamefont {Hills}}, \bibinfo {author}
  {\bibfnamefont {R.~P.}\ \bibnamefont {Campion}}, \bibinfo {author}
  {\bibfnamefont {V.}~\bibnamefont {Nov{\'a}k}}, \bibinfo {author}
  {\bibfnamefont {K.}~\bibnamefont {Olejn{\'\i}k}}, \bibinfo {author}
  {\bibfnamefont {F.}~\bibnamefont {Maccherozzi}}, \bibinfo {author}
  {\bibfnamefont {S.~S.}\ \bibnamefont {Dhesi}},  \emph {et~al.},\ }\href
  {https://science.sciencemag.org/content/351/6273/587} {\bibfield  {journal}
  {\bibinfo  {journal} {Science}\ }\textbf {\bibinfo {volume} {351}},\ \bibinfo
  {pages} {587} (\bibinfo {year} {2016})}\BibitemShut {NoStop}%
\bibitem [{\citenamefont {Olejn{\'\i}k}\ \emph {et~al.}(2017)\citenamefont
  {Olejn{\'\i}k}, \citenamefont {Schuler}, \citenamefont {Mart{\'\i}},
  \citenamefont {Nov{\'a}k}, \citenamefont {Ka{\v{s}}par}, \citenamefont
  {Wadley}, \citenamefont {Campion}, \citenamefont {Edmonds}, \citenamefont
  {Gallagher}, \citenamefont {Garc{\'e}s} \emph
  {et~al.}}]{olejnik2017antiferromagnetic}%
  \BibitemOpen
  \bibfield  {author} {\bibinfo {author} {\bibfnamefont {K.}~\bibnamefont
  {Olejn{\'\i}k}}, \bibinfo {author} {\bibfnamefont {V.}~\bibnamefont
  {Schuler}}, \bibinfo {author} {\bibfnamefont {X.}~\bibnamefont {Mart{\'\i}}},
  \bibinfo {author} {\bibfnamefont {V.}~\bibnamefont {Nov{\'a}k}}, \bibinfo
  {author} {\bibfnamefont {Z.}~\bibnamefont {Ka{\v{s}}par}}, \bibinfo {author}
  {\bibfnamefont {P.}~\bibnamefont {Wadley}}, \bibinfo {author} {\bibfnamefont
  {R.~P.}\ \bibnamefont {Campion}}, \bibinfo {author} {\bibfnamefont {K.~W.}\
  \bibnamefont {Edmonds}}, \bibinfo {author} {\bibfnamefont {B.~L.}\
  \bibnamefont {Gallagher}}, \bibinfo {author} {\bibfnamefont {J.}~\bibnamefont
  {Garc{\'e}s}},  \emph {et~al.},\ }\href
  {https://www.nature.com/articles/ncomms15434} {\bibfield  {journal} {\bibinfo
   {journal} {Nat. Commun.}\ }\textbf {\bibinfo {volume} {8}},\ \bibinfo
  {pages} {15434} (\bibinfo {year} {2017})}\BibitemShut {NoStop}%
\bibitem [{\citenamefont {Grzybowski}\ \emph {et~al.}(2017)\citenamefont
  {Grzybowski}, \citenamefont {Wadley}, \citenamefont {Edmonds}, \citenamefont
  {Beardsley}, \citenamefont {Hills}, \citenamefont {Campion}, \citenamefont
  {Gallagher}, \citenamefont {Chauhan}, \citenamefont {Novak}, \citenamefont
  {Jungwirth} \emph {et~al.}}]{grzybowski2017imaging}%
  \BibitemOpen
  \bibfield  {author} {\bibinfo {author} {\bibfnamefont {M.}~\bibnamefont
  {Grzybowski}}, \bibinfo {author} {\bibfnamefont {P.}~\bibnamefont {Wadley}},
  \bibinfo {author} {\bibfnamefont {K.}~\bibnamefont {Edmonds}}, \bibinfo
  {author} {\bibfnamefont {R.}~\bibnamefont {Beardsley}}, \bibinfo {author}
  {\bibfnamefont {V.}~\bibnamefont {Hills}}, \bibinfo {author} {\bibfnamefont
  {R.}~\bibnamefont {Campion}}, \bibinfo {author} {\bibfnamefont
  {B.}~\bibnamefont {Gallagher}}, \bibinfo {author} {\bibfnamefont {J.~S.}\
  \bibnamefont {Chauhan}}, \bibinfo {author} {\bibfnamefont {V.}~\bibnamefont
  {Novak}}, \bibinfo {author} {\bibfnamefont {T.}~\bibnamefont {Jungwirth}},
  \emph {et~al.},\ }\href
  {https://journals.aps.org/prl/abstract/10.1103/PhysRevLett.118.057701}
  {\bibfield  {journal} {\bibinfo  {journal} {Phys. Rev. Lett.}\ }\textbf
  {\bibinfo {volume} {118}},\ \bibinfo {pages} {057701} (\bibinfo {year}
  {2017})}\BibitemShut {NoStop}%
\bibitem [{\citenamefont {Zhou}\ \emph {et~al.}(2018)\citenamefont {Zhou},
  \citenamefont {Zhang}, \citenamefont {Li}, \citenamefont {Chen},
  \citenamefont {Shi}, \citenamefont {Tan}, \citenamefont {Gu}, \citenamefont
  {Saleem}, \citenamefont {Wu}, \citenamefont {Pan} \emph
  {et~al.}}]{zhou2018strong}%
  \BibitemOpen
  \bibfield  {author} {\bibinfo {author} {\bibfnamefont {X.}~\bibnamefont
  {Zhou}}, \bibinfo {author} {\bibfnamefont {J.}~\bibnamefont {Zhang}},
  \bibinfo {author} {\bibfnamefont {F.}~\bibnamefont {Li}}, \bibinfo {author}
  {\bibfnamefont {X.}~\bibnamefont {Chen}}, \bibinfo {author} {\bibfnamefont
  {G.}~\bibnamefont {Shi}}, \bibinfo {author} {\bibfnamefont {Y.}~\bibnamefont
  {Tan}}, \bibinfo {author} {\bibfnamefont {Y.}~\bibnamefont {Gu}}, \bibinfo
  {author} {\bibfnamefont {M.}~\bibnamefont {Saleem}}, \bibinfo {author}
  {\bibfnamefont {H.}~\bibnamefont {Wu}}, \bibinfo {author} {\bibfnamefont
  {F.}~\bibnamefont {Pan}},  \emph {et~al.},\ }\href
  {https://journals.aps.org/prapplied/abstract/10.1103/PhysRevApplied.9.054028}
  {\bibfield  {journal} {\bibinfo  {journal} {Phys. Rev. Appl.}\ }\textbf
  {\bibinfo {volume} {9}},\ \bibinfo {pages} {054028} (\bibinfo {year}
  {2018})}\BibitemShut {NoStop}%
\bibitem [{\citenamefont {Bodnar}\ \emph {et~al.}(2020)\citenamefont {Bodnar},
  \citenamefont {Skourski}, \citenamefont {Gomonay}, \citenamefont {Sinova},
  \citenamefont {Kl{\"a}ui},\ and\ \citenamefont
  {Jourdan}}]{bodnarmagnetoresistance2020}%
  \BibitemOpen
  \bibfield  {author} {\bibinfo {author} {\bibfnamefont {S.~Y.}\ \bibnamefont
  {Bodnar}}, \bibinfo {author} {\bibfnamefont {Y.}~\bibnamefont {Skourski}},
  \bibinfo {author} {\bibfnamefont {O.}~\bibnamefont {Gomonay}}, \bibinfo
  {author} {\bibfnamefont {J.}~\bibnamefont {Sinova}}, \bibinfo {author}
  {\bibfnamefont {M.}~\bibnamefont {Kl{\"a}ui}}, \ and\ \bibinfo {author}
  {\bibfnamefont {M.}~\bibnamefont {Jourdan}},\ }\href
  {https://journals.aps.org/prapplied/abstract/10.1103/PhysRevApplied.14.014004}
  {\bibfield  {journal} {\bibinfo  {journal} {Phys. Rev. Appl.}\ }\textbf
  {\bibinfo {volume} {14}},\ \bibinfo {pages} {014004} (\bibinfo {year}
  {2020})}\BibitemShut {NoStop}%
\bibitem [{\citenamefont {Barthem}\ \emph {et~al.}(2016)\citenamefont
  {Barthem}, \citenamefont {Colin}, \citenamefont {Haettel}, \citenamefont
  {Dufeu},\ and\ \citenamefont {Givord}}]{barthem2016easy}%
  \BibitemOpen
  \bibfield  {author} {\bibinfo {author} {\bibfnamefont {V.~M. T.~S.}\
  \bibnamefont {Barthem}}, \bibinfo {author} {\bibfnamefont {C.~V.}\
  \bibnamefont {Colin}}, \bibinfo {author} {\bibfnamefont {R.}~\bibnamefont
  {Haettel}}, \bibinfo {author} {\bibfnamefont {D.}~\bibnamefont {Dufeu}}, \
  and\ \bibinfo {author} {\bibfnamefont {D.}~\bibnamefont {Givord}},\ }\href
  {https://www.sciencedirect.com/science/article/abs/pii/S030488531530408X}
  {\bibfield  {journal} {\bibinfo  {journal} {J. Magn. Mat.}\ }\textbf
  {\bibinfo {volume} {406}},\ \bibinfo {pages} {289} (\bibinfo {year}
  {2016})}\BibitemShut {NoStop}%
\bibitem [{\citenamefont {Sapozhnik}\ \emph {et~al.}(2018)\citenamefont
  {Sapozhnik}, \citenamefont {Filianina}, \citenamefont {Bodnar}, \citenamefont
  {Lamirand}, \citenamefont {Mawass}, \citenamefont {Skourski}, \citenamefont
  {Elmers}, \citenamefont {Zabel}, \citenamefont {Kl{\"a}ui},\ and\
  \citenamefont {Jourdan}}]{sapozhnik2018direct}%
  \BibitemOpen
  \bibfield  {author} {\bibinfo {author} {\bibfnamefont {A.~A.}\ \bibnamefont
  {Sapozhnik}}, \bibinfo {author} {\bibfnamefont {M.}~\bibnamefont
  {Filianina}}, \bibinfo {author} {\bibfnamefont {S.~Y.}\ \bibnamefont
  {Bodnar}}, \bibinfo {author} {\bibfnamefont {A.}~\bibnamefont {Lamirand}},
  \bibinfo {author} {\bibfnamefont {M.-A.}\ \bibnamefont {Mawass}}, \bibinfo
  {author} {\bibfnamefont {Y.}~\bibnamefont {Skourski}}, \bibinfo {author}
  {\bibfnamefont {H.-J.}\ \bibnamefont {Elmers}}, \bibinfo {author}
  {\bibfnamefont {H.}~\bibnamefont {Zabel}}, \bibinfo {author} {\bibfnamefont
  {M.}~\bibnamefont {Kl{\"a}ui}}, \ and\ \bibinfo {author} {\bibfnamefont
  {M.}~\bibnamefont {Jourdan}},\ }\href
  {https://journals.aps.org/prb/abstract/10.1103/PhysRevB.97.134429} {\bibfield
   {journal} {\bibinfo  {journal} {Phys. Rev. B}\ }\textbf {\bibinfo {volume}
  {97}},\ \bibinfo {pages} {134429} (\bibinfo {year} {2018})}\BibitemShut
  {NoStop}%
\bibitem [{\citenamefont {Ka{\v{s}}par}\ \emph {et~al.}(2021)\citenamefont
  {Ka{\v{s}}par}, \citenamefont {Sur{\`y}nek}, \citenamefont {Zub{\'a}{\v{c}}},
  \citenamefont {Krizek}, \citenamefont {Nov{\'a}k}, \citenamefont {Campion},
  \citenamefont {Woernle}, \citenamefont {Gambardella}, \citenamefont {Marti},
  \citenamefont {N{\v{e}}mec} \emph {et~al.}}]{kavspar2021quenching}%
  \BibitemOpen
  \bibfield  {author} {\bibinfo {author} {\bibfnamefont {Z.}~\bibnamefont
  {Ka{\v{s}}par}}, \bibinfo {author} {\bibfnamefont {M.}~\bibnamefont
  {Sur{\`y}nek}}, \bibinfo {author} {\bibfnamefont {J.}~\bibnamefont
  {Zub{\'a}{\v{c}}}}, \bibinfo {author} {\bibfnamefont {F.}~\bibnamefont
  {Krizek}}, \bibinfo {author} {\bibfnamefont {V.}~\bibnamefont {Nov{\'a}k}},
  \bibinfo {author} {\bibfnamefont {R.~P.}\ \bibnamefont {Campion}}, \bibinfo
  {author} {\bibfnamefont {M.~S.}\ \bibnamefont {Woernle}}, \bibinfo {author}
  {\bibfnamefont {P.}~\bibnamefont {Gambardella}}, \bibinfo {author}
  {\bibfnamefont {X.}~\bibnamefont {Marti}}, \bibinfo {author} {\bibfnamefont
  {P.}~\bibnamefont {N{\v{e}}mec}},  \emph {et~al.},\ }\href
  {https://www.nature.com/articles/s41928-020-00506-4} {\bibfield  {journal}
  {\bibinfo  {journal} {Nat. Electr.}\ }\textbf {\bibinfo {volume} {4}},\
  \bibinfo {pages} {30} (\bibinfo {year} {2021})}\BibitemShut {NoStop}%
\bibitem [{\citenamefont {Janda}\ \emph {et~al.}(2020)\citenamefont {Janda},
  \citenamefont {Godinho}, \citenamefont {Ostatnicky}, \citenamefont
  {Pfitzner}, \citenamefont {Ulrich}, \citenamefont {Hoehl}, \citenamefont
  {Reimers}, \citenamefont {{\v{S}}ob{\'a}{\v{n}}}, \citenamefont {Metzger},
  \citenamefont {Reichlova} \emph {et~al.}}]{janda2020magneto}%
  \BibitemOpen
  \bibfield  {author} {\bibinfo {author} {\bibfnamefont {T.}~\bibnamefont
  {Janda}}, \bibinfo {author} {\bibfnamefont {J.}~\bibnamefont {Godinho}},
  \bibinfo {author} {\bibfnamefont {T.}~\bibnamefont {Ostatnicky}}, \bibinfo
  {author} {\bibfnamefont {E.}~\bibnamefont {Pfitzner}}, \bibinfo {author}
  {\bibfnamefont {G.}~\bibnamefont {Ulrich}}, \bibinfo {author} {\bibfnamefont
  {A.}~\bibnamefont {Hoehl}}, \bibinfo {author} {\bibfnamefont
  {S.}~\bibnamefont {Reimers}}, \bibinfo {author} {\bibfnamefont
  {Z.}~\bibnamefont {{\v{S}}ob{\'a}{\v{n}}}}, \bibinfo {author} {\bibfnamefont
  {T.}~\bibnamefont {Metzger}}, \bibinfo {author} {\bibfnamefont
  {H.}~\bibnamefont {Reichlova}},  \emph {et~al.},\ }\href
  {https://journals.aps.org/prmaterials/abstract/10.1103/PhysRevMaterials.4.094413}
  {\bibfield  {journal} {\bibinfo  {journal} {Phys. Rev. Mat.}\ }\textbf
  {\bibinfo {volume} {4}},\ \bibinfo {pages} {094413} (\bibinfo {year}
  {2020})}\BibitemShut {NoStop}%
\bibitem [{\citenamefont {Otxoa}\ \emph
  {et~al.}(2020{\natexlab{b}})\citenamefont {Otxoa}, \citenamefont {Atxitia},
  \citenamefont {Roy},\ and\ \citenamefont
  {Chubykalo-Fesenko}}]{otxoa2020giant}%
  \BibitemOpen
  \bibfield  {author} {\bibinfo {author} {\bibfnamefont {R.~M.}\ \bibnamefont
  {Otxoa}}, \bibinfo {author} {\bibfnamefont {U.}~\bibnamefont {Atxitia}},
  \bibinfo {author} {\bibfnamefont {P.~E.}\ \bibnamefont {Roy}}, \ and\
  \bibinfo {author} {\bibfnamefont {O.}~\bibnamefont {Chubykalo-Fesenko}},\
  }\href {https://www.nature.com/articles/s42005-020-0296-4} {\bibfield
  {journal} {\bibinfo  {journal} {Commun. Phys.}\ }\textbf {\bibinfo {volume}
  {3}},\ \bibinfo {pages} {31} (\bibinfo {year}
  {2020}{\natexlab{b}})}\BibitemShut {NoStop}%
\bibitem [{\citenamefont {Yang}\ \emph {et~al.}(2015)\citenamefont {Yang},
  \citenamefont {Ryu},\ and\ \citenamefont {Parkin}}]{yang2015domain}%
  \BibitemOpen
  \bibfield  {author} {\bibinfo {author} {\bibfnamefont {S.-H.}\ \bibnamefont
  {Yang}}, \bibinfo {author} {\bibfnamefont {K.-S.}\ \bibnamefont {Ryu}}, \
  and\ \bibinfo {author} {\bibfnamefont {S.}~\bibnamefont {Parkin}},\ }\href
  {https://www.nature.com/articles/nnano.2014.324} {\bibfield  {journal}
  {\bibinfo  {journal} {Nat. Nanotechnol.}\ }\textbf {\bibinfo {volume} {10}},\
  \bibinfo {pages} {221} (\bibinfo {year} {2015})}\BibitemShut {NoStop}%
\bibitem [{\citenamefont {Lequeux}\ \emph {et~al.}(2016)\citenamefont
  {Lequeux}, \citenamefont {Sampaio}, \citenamefont {Cros}, \citenamefont
  {Yakushiji}, \citenamefont {Fukushima}, \citenamefont {Matsumoto},
  \citenamefont {Kubota}, \citenamefont {Yuasa},\ and\ \citenamefont
  {Grollier}}]{lequeux2016magnetic}%
  \BibitemOpen
  \bibfield  {author} {\bibinfo {author} {\bibfnamefont {S.}~\bibnamefont
  {Lequeux}}, \bibinfo {author} {\bibfnamefont {J.}~\bibnamefont {Sampaio}},
  \bibinfo {author} {\bibfnamefont {V.}~\bibnamefont {Cros}}, \bibinfo {author}
  {\bibfnamefont {K.}~\bibnamefont {Yakushiji}}, \bibinfo {author}
  {\bibfnamefont {A.}~\bibnamefont {Fukushima}}, \bibinfo {author}
  {\bibfnamefont {R.}~\bibnamefont {Matsumoto}}, \bibinfo {author}
  {\bibfnamefont {H.}~\bibnamefont {Kubota}}, \bibinfo {author} {\bibfnamefont
  {S.}~\bibnamefont {Yuasa}}, \ and\ \bibinfo {author} {\bibfnamefont
  {J.}~\bibnamefont {Grollier}},\ }\href
  {https://www.nature.com/articles/srep31510} {\bibfield  {journal} {\bibinfo
  {journal} {Sci. Rep.}\ }\textbf {\bibinfo {volume} {6}},\ \bibinfo {pages}
  {1} (\bibinfo {year} {2016})}\BibitemShut {NoStop}%
\bibitem [{\citenamefont {Luo}\ \emph {et~al.}(2020)\citenamefont {Luo},
  \citenamefont {Hrabec}, \citenamefont {Dao}, \citenamefont {Sala},
  \citenamefont {Finizio}, \citenamefont {Feng}, \citenamefont {Mayr},
  \citenamefont {Raabe}, \citenamefont {Gambardella},\ and\ \citenamefont
  {Heyderman}}]{luo2020current}%
  \BibitemOpen
  \bibfield  {author} {\bibinfo {author} {\bibfnamefont {Z.}~\bibnamefont
  {Luo}}, \bibinfo {author} {\bibfnamefont {A.}~\bibnamefont {Hrabec}},
  \bibinfo {author} {\bibfnamefont {T.~P.}\ \bibnamefont {Dao}}, \bibinfo
  {author} {\bibfnamefont {G.}~\bibnamefont {Sala}}, \bibinfo {author}
  {\bibfnamefont {S.}~\bibnamefont {Finizio}}, \bibinfo {author} {\bibfnamefont
  {J.}~\bibnamefont {Feng}}, \bibinfo {author} {\bibfnamefont {S.}~\bibnamefont
  {Mayr}}, \bibinfo {author} {\bibfnamefont {J.}~\bibnamefont {Raabe}},
  \bibinfo {author} {\bibfnamefont {P.}~\bibnamefont {Gambardella}}, \ and\
  \bibinfo {author} {\bibfnamefont {L.~J.}\ \bibnamefont {Heyderman}},\ }\href
  {https://www.nature.com/articles/s41586-020-2061-y} {\bibfield  {journal}
  {\bibinfo  {journal} {Nature}\ }\textbf {\bibinfo {volume} {579}},\ \bibinfo
  {pages} {214} (\bibinfo {year} {2020})}\BibitemShut {NoStop}%
\bibitem [{\citenamefont {Shick}\ \emph {et~al.}(2010)\citenamefont {Shick},
  \citenamefont {Khmelevskyi}, \citenamefont {Mryasov}, \citenamefont
  {Wunderlich},\ and\ \citenamefont {Jungwirth}}]{shick2010spin}%
  \BibitemOpen
  \bibfield  {author} {\bibinfo {author} {\bibfnamefont {A.~B.}\ \bibnamefont
  {Shick}}, \bibinfo {author} {\bibfnamefont {S.}~\bibnamefont {Khmelevskyi}},
  \bibinfo {author} {\bibfnamefont {O.~N.}\ \bibnamefont {Mryasov}}, \bibinfo
  {author} {\bibfnamefont {J.}~\bibnamefont {Wunderlich}}, \ and\ \bibinfo
  {author} {\bibfnamefont {T.}~\bibnamefont {Jungwirth}},\ }\href
  {https://journals.aps.org/prb/abstract/10.1103/PhysRevB.81.212409} {\bibfield
   {journal} {\bibinfo  {journal} {Phys. Rev. B}\ }\textbf {\bibinfo {volume}
  {81}},\ \bibinfo {pages} {212409} (\bibinfo {year} {2010})}\BibitemShut
  {NoStop}%
\bibitem [{\citenamefont {Barthem}\ \emph {et~al.}(2013)\citenamefont
  {Barthem}, \citenamefont {Colin}, \citenamefont {Mayaffre}, \citenamefont
  {Julien},\ and\ \citenamefont {Givord}}]{barthem2013revealing}%
  \BibitemOpen
  \bibfield  {author} {\bibinfo {author} {\bibfnamefont {V.~M. T.~S.}\
  \bibnamefont {Barthem}}, \bibinfo {author} {\bibfnamefont {C.~V.}\
  \bibnamefont {Colin}}, \bibinfo {author} {\bibfnamefont {H.}~\bibnamefont
  {Mayaffre}}, \bibinfo {author} {\bibfnamefont {M.-H.}\ \bibnamefont
  {Julien}}, \ and\ \bibinfo {author} {\bibfnamefont {D.}~\bibnamefont
  {Givord}},\ }\href {https://www.nature.com/articles/ncomms3892} {\bibfield
  {journal} {\bibinfo  {journal} {Nat. Commun.}\ }\textbf {\bibinfo {volume}
  {4}},\ \bibinfo {pages} {2892} (\bibinfo {year} {2013})}\BibitemShut
  {NoStop}%
\bibitem [{\citenamefont {Masrour}\ \emph {et~al.}(2015)\citenamefont
  {Masrour}, \citenamefont {Hlil}, \citenamefont {Hamedoun}, \citenamefont
  {Benyoussef}, \citenamefont {Boutahar},\ and\ \citenamefont
  {Lassri}}]{masrour2015antiferromagnetic}%
  \BibitemOpen
  \bibfield  {author} {\bibinfo {author} {\bibfnamefont {R.}~\bibnamefont
  {Masrour}}, \bibinfo {author} {\bibfnamefont {E.~K.}\ \bibnamefont {Hlil}},
  \bibinfo {author} {\bibfnamefont {M.}~\bibnamefont {Hamedoun}}, \bibinfo
  {author} {\bibfnamefont {A.}~\bibnamefont {Benyoussef}}, \bibinfo {author}
  {\bibfnamefont {A.}~\bibnamefont {Boutahar}}, \ and\ \bibinfo {author}
  {\bibfnamefont {H.}~\bibnamefont {Lassri}},\ }\href
  {https://www.sciencedirect.com/science/article/abs/pii/S0304885315302092}
  {\bibfield  {journal} {\bibinfo  {journal} {J. Magn. Mat.}\ }\textbf
  {\bibinfo {volume} {393}},\ \bibinfo {pages} {600} (\bibinfo {year}
  {2015})}\BibitemShut {NoStop}%
\bibitem [{\citenamefont {Khmelevskyi}\ and\ \citenamefont
  {Mohn}(2008)}]{khmelevskyi2008layered}%
  \BibitemOpen
  \bibfield  {author} {\bibinfo {author} {\bibfnamefont {S.}~\bibnamefont
  {Khmelevskyi}}\ and\ \bibinfo {author} {\bibfnamefont {P.}~\bibnamefont
  {Mohn}},\ }\href {https://aip.scitation.org/doi/10.1063/1.3003878} {\bibfield
   {journal} {\bibinfo  {journal} {Appl. Phys. Lett.}\ }\textbf {\bibinfo
  {volume} {93}},\ \bibinfo {pages} {162503} (\bibinfo {year}
  {2008})}\BibitemShut {NoStop}%
\bibitem [{\citenamefont {Wells}\ and\ \citenamefont
  {Smith}(1970)}]{wells1970structure}%
  \BibitemOpen
  \bibfield  {author} {\bibinfo {author} {\bibfnamefont {P.}~\bibnamefont
  {Wells}}\ and\ \bibinfo {author} {\bibfnamefont {J.~H.}\ \bibnamefont
  {Smith}},\ }\href
  {https://onlinelibrary.wiley.com/doi/abs/10.1107/S056773947000092X}
  {\bibfield  {journal} {\bibinfo  {journal} {Acta Crystallogr. Sect. A}\
  }\textbf {\bibinfo {volume} {26}},\ \bibinfo {pages} {379} (\bibinfo {year}
  {1970})}\BibitemShut {NoStop}%
\bibitem [{\citenamefont {Roy}\ \emph {et~al.}(2016)\citenamefont {Roy},
  \citenamefont {Otxoa},\ and\ \citenamefont {Wunderlich}}]{roy2016robust}%
  \BibitemOpen
  \bibfield  {author} {\bibinfo {author} {\bibfnamefont {P.~E.}\ \bibnamefont
  {Roy}}, \bibinfo {author} {\bibfnamefont {R.~M.}\ \bibnamefont {Otxoa}}, \
  and\ \bibinfo {author} {\bibfnamefont {J.}~\bibnamefont {Wunderlich}},\
  }\href {https://journals.aps.org/prb/abstract/10.1103/PhysRevB.94.014439}
  {\bibfield  {journal} {\bibinfo  {journal} {Phys. Rev. B}\ }\textbf {\bibinfo
  {volume} {94}},\ \bibinfo {pages} {014439} (\bibinfo {year}
  {2016})}\BibitemShut {NoStop}%
\bibitem [{\citenamefont {Turov}\ \emph {et~al.}(2010)\citenamefont {Turov},
  \citenamefont {Kolchanov}, \citenamefont {Men'shenin}, \citenamefont
  {Mirsaev},\ and\ \citenamefont {Nikolaev}}]{turov2010symmetry}%
  \BibitemOpen
  \bibfield  {author} {\bibinfo {author} {\bibfnamefont {E.~A.}\ \bibnamefont
  {Turov}}, \bibinfo {author} {\bibfnamefont {A.~V.}\ \bibnamefont
  {Kolchanov}}, \bibinfo {author} {\bibfnamefont {V.~V.}\ \bibnamefont
  {Men'shenin}}, \bibinfo {author} {\bibfnamefont {I.~F.}\ \bibnamefont
  {Mirsaev}}, \ and\ \bibinfo {author} {\bibfnamefont {V.~V.}\ \bibnamefont
  {Nikolaev}},\ }\href@noop {} {\bibfield  {journal} {\bibinfo  {journal}
  {\textit{Symmetry and Physical Properties of Antiferromagnets}}\ } (\bibinfo
  {year} {Cambridge International Science Publishing, Cambridge,
  2010})}\BibitemShut {NoStop}%
\bibitem [{\citenamefont {Zvezdin}(1979)}]{zvezdin1979pis}%
  \BibitemOpen
  \bibfield  {author} {\bibinfo {author} {\bibfnamefont {A.~K.}\ \bibnamefont
  {Zvezdin}},\ }\href {http://jetpletters.ru/ps/1456/article_22180.shtml}
  {\bibfield  {journal} {\bibinfo  {journal} {JETP Lett.}\ }\textbf {\bibinfo
  {volume} {29}},\ \bibinfo {pages} {553} (\bibinfo {year} {1979})}\BibitemShut
  {NoStop}%
\bibitem [{\citenamefont {Bar'yakhtar}\ \emph {et~al.}(1985)\citenamefont
  {Bar'yakhtar}, \citenamefont {Ivanov},\ and\ \citenamefont
  {Chetkin}}]{bar1985dynamics}%
  \BibitemOpen
  \bibfield  {author} {\bibinfo {author} {\bibfnamefont {V.~G.}\ \bibnamefont
  {Bar'yakhtar}}, \bibinfo {author} {\bibfnamefont {B.~A.}\ \bibnamefont
  {Ivanov}}, \ and\ \bibinfo {author} {\bibfnamefont {M.~V.}\ \bibnamefont
  {Chetkin}},\ }\href
  {http://mr.crossref.org/iPage?doi=10.1070%2FPU1985v028n07ABEH003871}
  {\bibfield  {journal} {\bibinfo  {journal} {Sov. Phys. Uspekhi}\ }\textbf
  {\bibinfo {volume} {28}},\ \bibinfo {pages} {563} (\bibinfo {year}
  {1985})}\BibitemShut {NoStop}%
\bibitem [{\citenamefont {Lifshitz}\ and\ \citenamefont
  {Pitaevskii}(1980)}]{lifshitz1980statistical}%
  \BibitemOpen
  \bibfield  {author} {\bibinfo {author} {\bibfnamefont {E.~M.}\ \bibnamefont
  {Lifshitz}}\ and\ \bibinfo {author} {\bibfnamefont {L.~P.}\ \bibnamefont
  {Pitaevskii}},\ }\href
  {https://www.elsevier.com/books/statistical-physics/lifshitz/978-0-08-050350-9}
  {\bibfield  {journal} {\bibinfo  {journal} {\textit{Statistical Physics,
  Course of Theoretical Physics, Vol. 9}}\ } (\bibinfo {year} {Pergamon Press,
  Oxford, 1980})}\BibitemShut {NoStop}%
\bibitem [{\citenamefont
  {Papanicolaou}(1995)}]{papanicolaou1995antiferromagnetic}%
  \BibitemOpen
  \bibfield  {author} {\bibinfo {author} {\bibfnamefont {N.}~\bibnamefont
  {Papanicolaou}},\ }\href
  {https://journals.aps.org/prb/abstract/10.1103/PhysRevB.51.15062} {\bibfield
  {journal} {\bibinfo  {journal} {Phys. Rev. B}\ }\textbf {\bibinfo {volume}
  {51}},\ \bibinfo {pages} {15062} (\bibinfo {year} {1995})}\BibitemShut
  {NoStop}%
\bibitem [{\citenamefont {Tveten}\ \emph {et~al.}(2016)\citenamefont {Tveten},
  \citenamefont {M{\"u}ller}, \citenamefont {Linder},\ and\ \citenamefont
  {Brataas}}]{tveten2016intrinsic}%
  \BibitemOpen
  \bibfield  {author} {\bibinfo {author} {\bibfnamefont {E.~G.}\ \bibnamefont
  {Tveten}}, \bibinfo {author} {\bibfnamefont {T.}~\bibnamefont {M{\"u}ller}},
  \bibinfo {author} {\bibfnamefont {J.}~\bibnamefont {Linder}}, \ and\ \bibinfo
  {author} {\bibfnamefont {A.}~\bibnamefont {Brataas}},\ }\href
  {https://journals.aps.org/prb/abstract/10.1103/PhysRevB.93.104408} {\bibfield
   {journal} {\bibinfo  {journal} {Phys. Rev. B}\ }\textbf {\bibinfo {volume}
  {93}},\ \bibinfo {pages} {104408} (\bibinfo {year} {2016})}\BibitemShut
  {NoStop}%
\bibitem [{\citenamefont {Kosevich}\ \emph {et~al.}(1990)\citenamefont
  {Kosevich}, \citenamefont {Ivanov},\ and\ \citenamefont
  {Kovalev}}]{kosevich1990magnetic}%
  \BibitemOpen
  \bibfield  {author} {\bibinfo {author} {\bibfnamefont {A.~M.}\ \bibnamefont
  {Kosevich}}, \bibinfo {author} {\bibfnamefont {B.~A.}\ \bibnamefont
  {Ivanov}}, \ and\ \bibinfo {author} {\bibfnamefont {A.~S.}\ \bibnamefont
  {Kovalev}},\ }\href
  {https://www.sciencedirect.com/science/article/abs/pii/037015739090130T}
  {\bibfield  {journal} {\bibinfo  {journal} {Phys. Rep.}\ }\textbf {\bibinfo
  {volume} {194}},\ \bibinfo {pages} {117} (\bibinfo {year}
  {1990})}\BibitemShut {NoStop}%
\bibitem [{\citenamefont {Zvezdin}\ and\ \citenamefont
  {Kostyuchenko}(1999)}]{zvezdin1999nonlinear}%
  \BibitemOpen
  \bibfield  {author} {\bibinfo {author} {\bibfnamefont {A.~K.}\ \bibnamefont
  {Zvezdin}}\ and\ \bibinfo {author} {\bibfnamefont {V.~V.}\ \bibnamefont
  {Kostyuchenko}},\ }\href {https://link.springer.com/article/10.1134/1.559035}
  {\bibfield  {journal} {\bibinfo  {journal} {J. Exp. Theor. Phys.}\ }\textbf
  {\bibinfo {volume} {89}},\ \bibinfo {pages} {734} (\bibinfo {year}
  {1999})}\BibitemShut {NoStop}%
\bibitem [{\citenamefont {Arana}\ \emph {et~al.}(2017)\citenamefont {Arana},
  \citenamefont {Estrada}, \citenamefont {Maior}, \citenamefont {Mendes},
  \citenamefont {Fernandez-Outon}, \citenamefont {Macedo}, \citenamefont
  {Barthem}, \citenamefont {Givord}, \citenamefont {Azevedo},\ and\
  \citenamefont {Rezende}}]{arana2017observation}%
  \BibitemOpen
  \bibfield  {author} {\bibinfo {author} {\bibfnamefont {M.}~\bibnamefont
  {Arana}}, \bibinfo {author} {\bibfnamefont {F.}~\bibnamefont {Estrada}},
  \bibinfo {author} {\bibfnamefont {D.~S.}\ \bibnamefont {Maior}}, \bibinfo
  {author} {\bibfnamefont {J.~B.~S.}\ \bibnamefont {Mendes}}, \bibinfo {author}
  {\bibfnamefont {L.~E.}\ \bibnamefont {Fernandez-Outon}}, \bibinfo {author}
  {\bibfnamefont {W.~A.~A.}\ \bibnamefont {Macedo}}, \bibinfo {author}
  {\bibfnamefont {V.~M. T.~S.}\ \bibnamefont {Barthem}}, \bibinfo {author}
  {\bibfnamefont {D.}~\bibnamefont {Givord}}, \bibinfo {author} {\bibfnamefont
  {A.}~\bibnamefont {Azevedo}}, \ and\ \bibinfo {author} {\bibfnamefont
  {S.~M.}\ \bibnamefont {Rezende}},\ }\href
  {https://aip.scitation.org/doi/full/10.1063/1.5001705?casa_token=Pau3KX7MuCkAAAAA%3A4iUva7X9G6RAc_ytawwv3EXGyKQMzr8srQ8D96dKTGF9XV1CU_kRsDGJYFwwIzX-nIo8fwis2bQ}
  {\bibfield  {journal} {\bibinfo  {journal} {Appl. Phys. Lett.}\ }\textbf
  {\bibinfo {volume} {111}},\ \bibinfo {pages} {192409} (\bibinfo {year}
  {2017})}\BibitemShut {NoStop}%
\bibitem [{\citenamefont {Bar'yakhtar}\ and\ \citenamefont
  {Ivanov}(1980)}]{bar1980nonlinear}%
  \BibitemOpen
  \bibfield  {author} {\bibinfo {author} {\bibfnamefont {I.~V.}\ \bibnamefont
  {Bar'yakhtar}}\ and\ \bibinfo {author} {\bibfnamefont {B.~A.}\ \bibnamefont
  {Ivanov}},\ }\href
  {https://www.sciencedirect.com/science/article/abs/pii/0038109880901489?showall%3Dtrue}
  {\bibfield  {journal} {\bibinfo  {journal} {Solid State Commun.}\ }\textbf
  {\bibinfo {volume} {34}},\ \bibinfo {pages} {545} (\bibinfo {year}
  {1980})}\BibitemShut {NoStop}%
\bibitem [{\citenamefont {Andreev}\ and\ \citenamefont
  {Marchenko}(1980)}]{andreev1980symmetry}%
  \BibitemOpen
  \bibfield  {author} {\bibinfo {author} {\bibfnamefont {A.~F.}\ \bibnamefont
  {Andreev}}\ and\ \bibinfo {author} {\bibfnamefont {V.~I.}\ \bibnamefont
  {Marchenko}},\ }\href
  {https://iopscience.iop.org/article/10.1070/PU1980v023n01ABEH004859}
  {\bibfield  {journal} {\bibinfo  {journal} {Sov. Phys. Uspekhi}\ }\textbf
  {\bibinfo {volume} {23}},\ \bibinfo {pages} {21} (\bibinfo {year}
  {1980})}\BibitemShut {NoStop}%
\bibitem [{\citenamefont {Rajaraman}(1982)}]{rajaraman1982solitons}%
  \BibitemOpen
  \bibfield  {author} {\bibinfo {author} {\bibfnamefont {R.}~\bibnamefont
  {Rajaraman}},\ }\href
  {https://www.elsevier.com/books/solitons-and-instantons/rajaraman/978-0-444-87047-6}
  {\bibfield  {journal} {\bibinfo  {journal} {\textit{Solitons and
  Instantons}}\ } (\bibinfo {year} {North-Holland, Amsterdam,
  1982})}\BibitemShut {NoStop}%
\bibitem [{\citenamefont {Schryer}\ and\ \citenamefont
  {Walker}(1974)}]{schryer1974motion}%
  \BibitemOpen
  \bibfield  {author} {\bibinfo {author} {\bibfnamefont {N.~L.}\ \bibnamefont
  {Schryer}}\ and\ \bibinfo {author} {\bibfnamefont {L.~R.}\ \bibnamefont
  {Walker}},\ }\href {https://aip.scitation.org/doi/10.1063/1.1663252}
  {\bibfield  {journal} {\bibinfo  {journal} {J. Appl. Phys.}\ }\textbf
  {\bibinfo {volume} {45}},\ \bibinfo {pages} {5406} (\bibinfo {year}
  {1974})}\BibitemShut {NoStop}%
\bibitem [{\citenamefont {Gomonay}\ \emph {et~al.}(2018)\citenamefont
  {Gomonay}, \citenamefont {Jungwirth},\ and\ \citenamefont
  {Sinova}}]{gomonay2018narrow}%
  \BibitemOpen
  \bibfield  {author} {\bibinfo {author} {\bibfnamefont {O.}~\bibnamefont
  {Gomonay}}, \bibinfo {author} {\bibfnamefont {T.}~\bibnamefont {Jungwirth}},
  \ and\ \bibinfo {author} {\bibfnamefont {J.}~\bibnamefont {Sinova}},\ }\href
  {https://journals.aps.org/prb/abstract/10.1103/PhysRevB.98.104430} {\bibfield
   {journal} {\bibinfo  {journal} {Phys. Rev. B}\ }\textbf {\bibinfo {volume}
  {98}},\ \bibinfo {pages} {104430} (\bibinfo {year} {2018})}\BibitemShut
  {NoStop}%
\bibitem [{cla()}]{clarification}%
  \BibitemOpen
  \href@noop {} {}\bibinfo {note} {It should be noted that the absence of
  periodicity along the $z$-{\it th} spatial direction in the simulated
  ultrathin film does not guarantee the condition $\boldsymbol{m}_{1,2} =
  \boldsymbol{m}_{3,4}$ used in Section \ref{section:theory}, which allowed us
  to reduce the system description to a two-sublattice-based nonlinear
  $\sigma$-model.}\BibitemShut {Stop}%
\end{thebibliography}%

\end{document}


\title{Supplemental Material: ``Inertial domain wall characterization in layered multisublattice antiferromagnets''}

\author{R. Rama-Eiroa}
\affiliation{Donostia International Physics Center, 20018 San Sebasti\'an, Spain}
\affiliation{Polymers and Advanced Materials Department: Physics, Chemistry, and Technology, University of the Basque Country, UPV/EHU, 20018 San Sebasti\'an, Spain}
\author{P. E. Roy}
\affiliation{Hitachi Cambridge Laboratory, J. J. Thomson Avenue, Cambridge CB3 0HE, United Kingdom}
\author{J. M. Gonz\'alez}
\affiliation{Polymers and Advanced Materials Department: Physics, Chemistry, and Technology, University of the Basque Country, UPV/EHU, 20018 San Sebasti\'an, Spain}
\author{K. Y. Guslienko}
\affiliation{Polymers and Advanced Materials Department: Physics, Chemistry, and Technology, University of the Basque Country, UPV/EHU, 20018 San Sebasti\'an, Spain}
\affiliation{IKERBASQUE, the Basque Foundation for Science, Plaza Euskadi, 5,
48009 Bilbao, Spain}
\author{J. Wunderlich}
\affiliation{Institute of Physics ASCR, v.v.i., Cukrovarnicka 10, 162 53 Praha 6, Czech Republic}
\affiliation{Institute of Experimental and Applied Physics, University of Regensburg, Universit\"atsstra{\ss}e 31, 93051 Regensburg, Germany}
\author{R. M. Otxoa}
\affiliation{Hitachi Cambridge Laboratory, J. J. Thomson Avenue, Cambridge CB3 0HE, United Kingdom}
\affiliation{Donostia International Physics Center, 20018 San Sebasti\'an, Spain}

\maketitle

\section*{SUPPLEMENTARY NOTE I: ``SIMULATED EXCHANGE-BASED INTRINSIC DOMAIN WALL PROPERTIES''}\label{sec:J2}

\begin{figure*}[hbt]
\centering
\includegraphics[width=18cm]{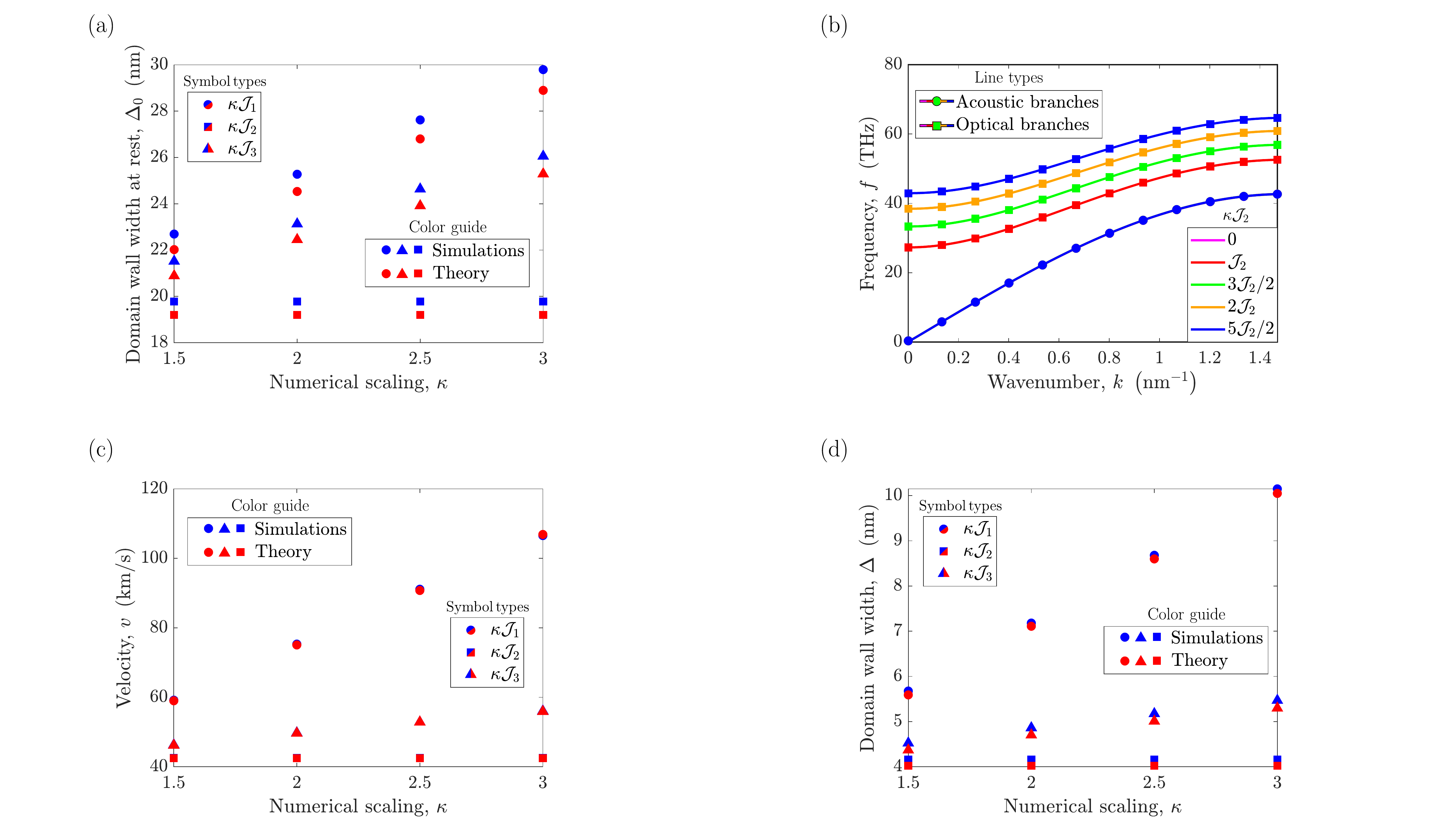}
\caption{(a) Comparison between the simulated and the theoretically predicted DW width at rest, $\Delta_0$, for different numerical scalings, $\kappa$, of, at each time, one of the exchange parameters of the system, $\mathcal{J}_1$, $\mathcal{J}_2$, and $\mathcal{J}_3$. (b) Simulated SW dispersion relations, including the high-frequency optical and low-frequency acoustic branches, characterized through the associated SW frequency $f$ and wavenumber $k$, for different numerical scaling through the $\kappa$ parameter of the AFM exchange interaction governed by $\mathcal{J}_2$. Comparison between the simulated and the analytically calculated steady-state velocity $v$ (c) and DW width $\Delta$ (d), for the case of a SO field of $H^{\mathrm{SO}}_y=60$ mT, for different numerical scalings, controlled by the $\kappa$ parameter, affecting, individually in each case, one of the exchange interactions of the system, characterized by $\mathcal{J}_1$, $\mathcal{J}_2$, and $\mathcal{J}_3$.}
\label{im:s1}
\end{figure*}

To elucidate the role that each exchange interaction plays in the static and dynamic domain wall (DW) properties, we have performed atomistic spin dynamics simulations where we have varied, through a numerical scaling factor, given by $\kappa$, the magnitude of the exchange parameters of the system. By varying only one of these at each time, we can compare the analytical expressions shown in Section III of the main text with the results that we obtain from the simulations. First, and as it can be seen in Suppl. Fig. \ref{im:s1} (a), the simulated DW width at rest $\Delta_0$ does not depend on the scaling magnitude applied to the antiferromagnetic (AFM) $\mathcal{J}_2$ constant, since this interaction only guarantees that the contiguous sublattices are perfectly antiparallel, while $\mathcal{J}_1$ and $\mathcal{J}_3$ characterize the relative angle between in-plane spins and hence the spatial soliton extent. This is consistent with the expression that we have obtained in Section III of the main text, according to which the DW width at rest does not depend on the contribution governed by $\mathcal{J}_2$, that is, $\Delta_0 = a_0 \sqrt{\left( \mathcal{J}_3 + \abs{\mathcal{J}_1}/2 \right) /K_{2 \parallel}}$. As it can be appreciated, the correlation between the trend of the simulated and analytically-predicted DW width at rest $\Delta_0$ is clear. On the other hand, we can explore the spin wave (SW) dispersion relation of the system by varying only the magnitude of the $\mathcal{J}_2$ parameter, and see how this affects the functional form of the low-frequency acoustic branch from which it can be extracted the maximum magnon group velocity, $v_{\mathrm{m}}$, of the medium, which is characterized as $v_{\mathrm{m}}={\left( \diff f/ \diff k \right)}_{\mathrm{max}}$, where $f$ expresses the SW frequency and $k$ represents the associated wavenumber along the propagation direction. As it can be seen in Suppl. Fig. \ref{im:s1} (b), the low-frequency acoustic branch is independent of the magnitude associated with the exchange interaction governed by $\mathcal{J}_2$, while the high-frequency optical branch maintains its functional form for $ \kappa \mathcal{J}_2 \neq 0$, but it shifts to higher frequencies as the value of the aforementioned combined variable increases. When $\kappa=0$, the optical branch disappears, so it is strongly dependent on the magnitude of $\mathcal{J}_2$. However, the maximum slope of the differentiated SW dispersion relation is always found in the $\mathcal{J}_2$-independent acoustic branch, so the maximum magnon group velocity, $v_{\mathrm{m}}$, does not depend on the aforementioned interaction.

Now we can evaluate the dynamic behaviour of the magnetic texture under the action of a constant current-induced spin-orbit (SO) field, $H^{\mathrm{SO}}_y=60$ mT. Taking into account that, as we introduced in Eq. (8) in the main text, the steady-state speed, $v$, has a functional form such that $v=v \left( \Delta_0, v_{\mathrm{m}} \right)$, no dependency on $\mathcal{J}_2$ is expected. Accordingly, in Suppl. Fig. \ref{im:s1} (c) it is possible to see that there is a great correspondence between the simulated and analytically predicted values for the steady-state velocity $v$ for different numerical scalings in the exchange parameters $\mathcal{J}_1$, $\mathcal{J}_2$, and $\mathcal{J}_3$. While it does not depend on the intersublattice AFM exchange contribution $\mathcal{J}_2$, it is strongly dependent on the magnitude of the interlayer AFM exchange interaction $\mathcal{J}_1$, which is consistent with the analytic form proposed, and is affected by changes in the value of the intrasublattice ferromagnetic (FM) exchange term encoded by $\mathcal{J}_3$. On the other hand, because the theoretically-predicted steady-state DW width, $\Delta$, is entirely conditioned by its rest state value, $\Delta_0$, the corresponding speed $v$ to the applied SO field $H^{\mathrm{SO}}_y$, and the maximum magnon group velocity of the medium, $v_{\mathrm{m}}$, since $\Delta = \Delta_0 \, \beta$, being $\beta$ the Lorentz factor, again the AFM exchange interaction encoded through the $\mathcal{J}_2$ parameter should not play any role in its determination. Consequently, and as it can be seen in Suppl. Fig. \ref{im:s1} (d), the steady-state DW width follows the theoretically-predicted trend, only existing, as in the case of the DW width at rest $\Delta_0$ presented in Suppl. Fig. \ref{im:s1} (a), a small constant shifting between the simulated and the theoretical values. Thus, the role of $\mathcal{J}_2$ consists of acting as a hard-axis anisotropy, strongly constraining the magnetization in the Mn-based basal planes of the conventional unit cell, since this interaction is the dominant one in magnitude of the system, but it does not intervene in the static and dynamic DW configurational properties.

\section*{SUPPLEMENTARY NOTE II: ``EXCHANGE ENERGY-BASED ANALYTICAL DESCRIPTION''}\label{sec:exchange}

To construct the exchange-based part of the energy, $w$, specified in Eq. (2) of the main text, it is necessary to take into account the number of first nearest neighbours along the $x$-{\it th} spatial direction and through which exchange interactions are mediated. For this, we will rely on Fig. 1 (c) of the manuscript, and we will take as a reference an atom of the layer 2, according to the numbering of the sheets shown in Fig. 1 (a) of the main text, in an arbitrary position $x_i$, whose unit magnetization vector will be denoted as $\boldsymbol{m}_2 \left( x_i \right)$. Since, along the $x$-{\it th} axis, the intra- and interlayer first nereast neighbours are at different distances, we will spatially denote those of layer 1 through $x_{i \pm 1/2}$, characterizing the $\mathcal{J}_1$ interaction, and those of layer 2 through $x_{i \pm 1}$, mediated by the $\mathcal{J}_3$ coefficient. This being the case, taking into account that each atom of layer 2 has four first nearest neighbours from layer 1 at a distance $a_0/2$ along the $x$-{\it th} spatial direction, we can first characterize the $\mathcal{J}_1$ contribution to the exchange energy of the system through the introduction of the unit magnetization vectors $\boldsymbol{m}_1 \left( x_{i \pm 1/2} \right)$ and their interaction with the magnetization vector of the reference atom $\boldsymbol{m}_2 \left( x_i \right)$, that is
\begin{gather}
w_{\mathrm{exc}} \left( \mathcal{J}_1 \right)=-2 \mathcal{J}_1 \, \boldsymbol{m}_2 \left( x_i \right) \cdot \left[ \boldsymbol{m}_1 \left( x_{i-1/2} \right)+\boldsymbol{m}_1 \left( x_{i+1/2} \right) \right],
\tag{S1}
\label{eq:S1}
\end{gather}
where it is important to note that the $2$ factor that appears on the right-hand side corresponds to the fact that there are two first nearest neighbours on each side of the reference atom of the layer 2 located at $x_i$.

Performing a one-dimensional Taylor series decomposition up to second order for the atoms belonging to layer 1 with respect to the reference position $x_i$ of the layer 2 atom, we will have
\begin{gather}
w_{\mathrm{exc}} \left( \mathcal{J}_1 \right)=4 \abs{\mathcal{J}_1} \, \boldsymbol{m}_2 \left( x_i \right) \cdot \left[ \boldsymbol{m}_1 \left( x_i \right)+  \frac{1}{2} {\left( \frac{a_0}{2} \right)}^2 \, \left( \partial^2_x \boldsymbol{m}_1 \left( x_i \right) \right) \right],
\tag{S2}
\label{eq:S2}
\end{gather}
expression which can be rewritten in terms of the magnetization vector, $\boldsymbol{m} = \left( \boldsymbol{m}_{\mathrm{A}} + \boldsymbol{m}_{\mathrm{B}} \right) / 2 $, and the N\'eel order parameter, $ \boldsymbol{l} = \left( \boldsymbol{m}_{\mathrm{A}} - \boldsymbol{m}_{\mathrm{B}} \right) / 2 $, where, in this case, due to the equivalence between the magnetic sublattices 1-3, which we will denote as type A, as it can be seen in Fig. 1 (b) of the main text, and the correlation between the layers 2-4, being able to refer to them as type B, as explained in Section III of the manuscript, it is possible to redefine, in this case, the orthogonal set of unitary vectors $\boldsymbol{m}$ and $\boldsymbol{l}$, as $\boldsymbol{m} = \left( \boldsymbol{m}_1 + \boldsymbol{m}_2 \right) / 2$ and $\boldsymbol{l} = \left( \boldsymbol{m}_1 - \boldsymbol{m}_2 \right) / 2$. Considering that $\boldsymbol{m}^2 + \boldsymbol{l}^2 = 1$ and that we are working on the exchange limit regarding the exchange interaction $\mathcal{J}_1$ ($\boldsymbol{m} \ll \boldsymbol{l}$), it is possible to obtain that, after throwing out the spatial indices, the exchange interaction can be reduced to
\begin{equation}
w_{\mathrm{exc}} \left( \mathcal{J}_1 \right) = 8 \abs{\mathcal{J}_1} \left[ \boldsymbol{m}^2-\frac{1}{4} {\left( \frac{a_0}{2} \right)}^2 \, \boldsymbol{l} \cdot \left( \partial^2_x \boldsymbol{l} \right) \right].
\tag{S3}
\label{eq:S3}
\end{equation}

Finally, taking into account that the N\'eel order parameter $\boldsymbol{l}$, as well as the magnetization vector $\boldsymbol{m}$, is a fixed length vector, it is possible to demonstrate that $\boldsymbol{l} \cdot \left( \partial^2_x \boldsymbol{l} \right)=-{\left( \partial_x \boldsymbol{l} \right)}^2$. Hence the exchange contribution encoded via $\mathcal{J}_1$ can be rewritten as
\begin{equation}
w_{\mathrm{exc}} \left( \mathcal{J}_1 \right) = 8 \abs{\mathcal{J}_1} \left[ \boldsymbol{m}^2+\frac{a^2_0}{16}  \, {\left( \partial_x \boldsymbol{l} \right)}^2 \right].
\tag{S4}
\label{eq:S4}
\end{equation}

Similarly, the functional form of the intrasublattice FM exchange parameter $\mathcal{J}_3$ contribution to the system can be analyzed analogously to the case concerning layer 1. For this, the interaction of the unit magnetization vector of layer 2 in an arbitrary spatial position, $\boldsymbol{m}_2 \left( x_i \right)$, with its two first nearest neighbours along the $x$-{\it th} axis located at a distance $a_0$ within the same sublattice must be taken into account, which will be given by
\begin{gather}
w_{\mathrm{exc}} \left( \mathcal{J}_3 \right)=- \mathcal{J}_3 \, \boldsymbol{m}_2 \left( x_i \right) \cdot \left[ \boldsymbol{m}_2 \left( x_{i-1} \right)+\boldsymbol{m}_2 \left( x_{i+1} \right) \right].
\tag{S7}
\label{eq:S5}
\end{gather}

Performing a Taylor series expansion up to second order along the $x$-{\it th} spatial direction of the intrasublattice first nearest neighbours atoms at a distance $a_0$ from the reference atom of layer 2 located at the arbitrary position $x_i$, it is possible to get that
\begin{gather}
w_{\mathrm{exc}} \left( \mathcal{J}_3 \right)=-2 \mathcal{J}_3 \, \boldsymbol{m}_2 \left( x_i \right) \cdot \left[ \boldsymbol{m}_2 \left( x_i \right)+\frac{a^2_0}{2} \, \left( \partial^2_x \boldsymbol{m}_2 \left( x_i \right) \right) \right].
\tag{S8}
\label{eq:S6}
\end{gather}

It is possible to rewrite this equation in terms of the AFM $\boldsymbol{l}$ and the magnetization $\boldsymbol{m}$ vectors given by the sublattice categories denoted by labels A and B, as we did earlier in Eq. \eqref{eq:S3}, not being necessary to specify in this case which numerically-based layer of type A we are basing ourselves on to construct the vectors, since sublattices 1-3 are magnetically equivalent, as we pointed out previously. If we take into account, again, that we are working on the exchange limit, that the $\boldsymbol{m}$ and $\boldsymbol{l}$ variables form an orthogonal unit set ($\boldsymbol{m} \cdot \boldsymbol{l}=0$), and that both are fixed length vectors, we can get that
\begin{equation}
w_{\mathrm{exc}} \left( \mathcal{J}_3 \right) = \mathcal{J}_3 a^2_0 \, {\left( \partial_x \boldsymbol{l} \right)}^2.
\tag{S9}
\label{eq:S7}
\end{equation}

Finally, it is possible to group both contributions, given by Eqs. \eqref{eq:S4} and \eqref{eq:S7}, which will describe the static and dynamic DW configuration in one of the FM layers of the conventional tetragonal unit cell, giving rise to
\begin{equation}
w_{\mathrm{exc}}=\frac{1}{2} \, A \, {\boldsymbol{m}}^2+\frac{1}{8} \, a \, {\left( \partial_x \boldsymbol{l} \right)}^2,
\tag{S10}
\label{eq:S8}
\end{equation}
where we have introduced the homogeneous AFM exchange parameter, $A=16 \abs{\mathcal{J}_1}$, and the inhomogeneous FM-like exchange constant, given by $a=8 a^2_0 \left( \mathcal{J}_3+\abs{\mathcal{J}_1}/2 \right)$. It should be noted the absence of a potential contribution by the AFM exchange interaction encoded through the $\mathcal{J}_2$ parameter in the previous expression, in line with the analysis carried out in Suppl. Note I.